\documentclass{aa}  

\usepackage{graphicx}
\usepackage{txfonts}
\usepackage{textcomp}
\usepackage{multirow}
\usepackage[colorlinks=true,
            linkcolor=red,
            urlcolor=blue,
            citecolor=blue]{hyperref}

\usepackage{placeins}

\begin{document}
    
    \title{Large Interferometer For Exoplanets (\emph{LIFE}):}
    \subtitle{VI. Detecting rocky exoplanets in the habitable zones of Sun-like stars}
    \titlerunning{VI. Detecting rocky exoplanets in the habitable zones of Sun-like stars}

    \author{Jens~Kammerer\inst{\ref{stsci}}\fnmsep\thanks{Correspondence: \href{mailto:jkammerer@stsci.edu}{jkammerer@stsci.edu}.} \and
            Sascha~P.~Quanz\inst{\ref{eth},\ref{nccr}} \and
            Felix~Dannert\inst{\ref{eth}} \and
            the \emph{LIFE} Collaboration\inst{\ref{life}} 
            }
    \authorrunning{Kammerer et al.}
    
    \institute{Space Telescope Science Institute, 3700 San Martin Dr, Baltimore, MD 21218, USA\label{stsci}
    \and
    ETH Zurich, Institute for Particle Physics \& Astrophysics, Wolfgang-Pauli-Str. 27, 8093 Zurich, Switzerland\label{eth}
    \and
    National Center of Competence in Research PlanetS (\url{www.nccr-planets.ch})\label{nccr}
    \and
    \url{www.life-space-mission.com}\label{life}
    }

    \date{Received: <date> / Accepted: <date>}
    
    \abstract
    {Identifying and characterizing habitable and potentially inhabited worlds is one of the main goals of future exoplanet direct-imaging missions. The number of planets within the habitable zone (HZ) that are accessible to such missions is a key metric to quantify their scientific potential, and it can drive the mission and instrument design.}
    {While previous studies have shown a strong preference for a future mid-infrared nulling interferometer space mission, such as \emph{LIFE}, to detect planets within the HZ around M dwarfs, we here focus on a more conservative approach toward the concept of habitability and present yield estimates for two stellar samples consisting of nearby ($d < 20~\text{pc}$) Sun-like stars ($4800~\text{K} \leq T_\text{eff} \leq 6300~\text{K}$) and nearby FGK-type stars ($3940~\text{K} \leq T_\text{eff} \leq 7220~\text{K}$) accessible to such a mission.}
    {Our yield estimates are based on recently derived occurrence rates of rocky planets from the \emph{Kepler} mission and our \emph{LIFE} exoplanet observation simulation tool \texttt{LIFEsim}, which includes all main astrophysical noise sources, but no instrumental noise sources as yet. In a Monte Carlo-like approach, we marginalized over 1'000 synthetic planet populations simulated around single and wide binary stars from our two samples. We use new occurrence rates for rocky planets that cover the entire HZ around FGK-type stars, marginalize over the uncertainties in the underlying occurrence rate model, present a parameter study investigating the dependence of the planet yield on different instrumental and astrophysical parameters, and estimate the number of detectable HZ planets that might indeed harbor liquid surface water.}
    {Depending on a pessimistic or optimistic extrapolation of the \emph{Kepler} results, we find that during a 2.5-year search phase, \emph{LIFE} could detect between $\sim10$--16 (average) or $\sim5$--34 (including 1$\sigma$ uncertainties) rocky planets ($0.5~\text{R}_\oplus \leq R_\text{p} \leq 1.5~\text{R}_\oplus$) within the optimistic HZ of Sun-like stars and between $\sim4$--6 (average) or $\sim1$--13 (including 1$\sigma$ uncertainties) exo-Earth candidates (EECs) assuming four collector spacecraft equipped with 2~m mirrors and a conservative instrument throughput of 5\%. The error bars are dominated by uncertainties in the underlying planet occurrence rates and the extrapolation of the \emph{Kepler} results. With $D = 3.5~\text{m}$ or 1~m mirrors, the yield $Y$ changes strongly, following approximately $Y \propto D^{3/2}$. With the larger sample of FGK-type stars, the yield increases to $\sim16$--22 (average) rocky planets within the optimistic HZ and $\sim5$--8 (average) EECs, which corresponds to $\sim50\%$ of the yield predicted for M dwarfs in \emph{LIFE} paper I. Furthermore, we find that in addition to the mirror diameter, the yield depends strongly on the total throughput, but only weakly on the exozodiacal dust level and the accessible wavelength range of the mission.}
    {When the focus lies entirely on Sun-like stars, larger mirrors ($\sim3~\text{m}$ with 5\% total throughput) or a better total throughput ($\sim20\%$ with 2~m mirrors) are required to detect a statistically relevant sample of $\sim30$ rocky planets within the optimistic HZ. When the scope is extended to FGK-type stars, and especially when M dwarfs are included, a significant increase in the number of detectable rocky HZ planets is obtained, which relaxes the requirements on mirror size and total throughput. Observational insight into the habitability of planets orbiting M dwarfs, for example, from the \emph{James Webb Space Telescope}, is crucial for guiding the target selection and observing sequence optimization for a mission such as \emph{LIFE}.}
    
    
    \maketitle
    
    \section{Introduction}
    \label{sec:introduction}
    
    The search for habitable and even inhabited extrasolar worlds is arguably the most thrilling long-term goal of exoplanet science. Transit surveys have been successful in discovering a small sample of terrestrial planets that are roughly as large as the Earth and might retain atmospheres \citep[e.g.,][]{berta-thompson2015,gillon2017,vanderspek2019,gilbert2020}. In the near future, the \emph{Atmospheric Remote-sensing Infrared Exoplanet Large-survey} (\emph{Ariel}) space mission will provide a statistical census of exoplanet atmospheres down to super-Earth-sized planets \citep{tinetti2018}, and it is expected that the \emph{James Webb Space Telescope} (\emph{JWST}) will be able to probe the atmospheres of a handful of warm or temperate nearby transiting and Earth-sized planets such as those in the TRAPPIST-1 system \citep{lustig-yaeger2019,koll2019}. While these observations will revolutionize the study of exoplanet atmospheres, a comprehensive characterization including a detailed search for so-called biomarkers in these atmospheres is hardly possible with the \emph{JWST} \citep{barstow2016}. Ultimately, a more sensitive mission is therefore needed and a much larger sample of Earth-like planets needs to be discovered and investigated to place meaningful constraints on the occurrence of habitable worlds in the solar neighborhood and the likelihood of the existence of life on these worlds \citep[e.g.,][]{quanz2021_whitepaper}.
    
    To date, most known exoplanets have been discovered with transit and radial velocity techniques\footnote{\url{https://exoplanetarchive.ipac.caltech.edu/}} \citep[e.g.,][]{pepe2014}, which are both indirect detection techniques. While the atmospheres of transiting exoplanets can be exquisitely investigated with transit and secondary eclipse spectroscopy \citep[e.g.,][]{seager2010,deming2017,louie2018}, transiting planets that reside within the habitable zone (HZ) of their host star are statistically rare in the solar neighborhood. For example, \citet{bryson2021} estimated with 95\% confidence that there should only be $\text{about four}$ HZ planets around GK-type stars within $d = 10~\text{pc}$. Assuming an optimistic transit probability for HZ planets of 1\% and that the frequency of planets scales with $d^3$, this results in an expectation value of only $\sim0.32$ transiting HZ planets around GK-type stars within $d = 20~\text{pc}$. Moreover, transiting HZ planets are strongly biased toward M dwarfs \citep[e.g., TRAPPIST-1,][]{gillon2017} because the transit signal is stronger and the HZ is located closer to the host star, which increases the geometrical transit probability. While atmospheric characterization of (nontransiting) radial velocity planets is possible with the cross-correlation Doppler spectroscopy technique, this method requires very high spectral resolution \citep[$R \sim 100'000,$][]{dekok2014}, which is challenging for faint Earth-like exoplanets even with the upcoming Extremely Large Telescopes \citep[ELTs; e.g.,][]{snellen2015}.
    
    Direct detection techniques are therefore needed to efficiently discover and characterize Earth-like exoplanets in the solar neighborhood. From the ground, high-contrast instruments on the ELTs will be able to directly detect a small sample of rocky ($<2~\text{R}_\oplus$) exoplanets around some of the nearest stars \citep[e.g.,][]{quanz2015,bowens2021,kasper2021}. However, only next-generation space mission concepts such as \emph{LUVOIR} \citep[\emph{Large Ultraviolet Optical Infrared Telescope}][]{luvoir2019}, \emph{HabEx} \citep[\emph{Habitable Exoplanet Observatory}][]{gaudi2020}, and \emph{LIFE} \citep[\emph{Large Interferometer For Exoplanets}][]{quanz2021,quanz2021_whitepaper} will be able to directly detect and characterize a large sample of nearby terrestrial exoplanets. The recently published Astro2020 Decadal Survey on Astronomy and Astrophysics indeed recommends a $\sim6~\text{m}$ inscribed diameter infrared, optical, and ultraviolet space telescope aiming (among other science goals) at discovering and characterizing Earth-like exoplanets in reflected light around nearby Sun-like stars shortward of $\sim1~\text{\textmu m}$ as the next NASA flagship mission after the \emph{JWST} and the \emph{Nancy Roman Space Telescope}. In contrast, the European-led \emph{LIFE} mission concept aims for the characterization of Earth-like exoplanets in the mid-infrared (MIR) thermal light regime using a space-based nulling interferometer. The mid-infrared wavelength regime harbors a rich set of absorption features of important atmospheric molecules, including biosignatures, that could indicate the presence of life on an exoplanet \citep[e.g., ozone and methane;][]{schwieterman2018,catling2018} and furthermore provides direct constraints on the radius and atmospheric structure of the observed exoplanets \citep[e.g.,][]{konrad2021}. Both approaches, investigations in reflected light and in thermal emission, have their advantages and disadvantages. When they are combined, however, they provide the deepest insight into the habitability of alien worlds. It is currently unclear whether any of these missions will feature an initial search phase to first identify the most promising targets for an in-depth characterization in a subsequent mission phase or whether these targets can already be previously identified using large-scale radial velocity surveys and other techniques from the ground \citep[e.g.,][]{luvoir2019}. Nevertheless, the number of Earth-like exoplanets in the HZ of their host star that could be found in an initial search phase is an important metric for evaluating the scientific merit of all of these mission concepts and inform science and technical requirements.
    
    In previous works, \citet{stark2014} and \citet{stark2015} found that a reflected-light coronagraphy space mission needs an aperture of at least $\sim8.5~\text{m}$ circumscribed diameter to achieve an exo-Earth candidate (EEC) yield of 30. Recent studies for a thermal emission nulling interferometry space mission such as \emph{LIFE} by \citet{kammerer2018}, \citet{quanz2018}, and \citet{quanz2021} showed that when a total instrument throughput of 5\% is assumed, four collector spacecraft with $\sim3.5~\text{m}$ apertures are required to obtain a similar yield of 30 EECs. These yield estimates also revealed a strong preference for detections in thermal emission around M dwarfs. This stems from the high fraction of M dwarfs in the solar neighborhood \citep[e.g.,][]{kroupa2001,reyle2021} and the limiting factor of the nulling interferometry technique being sensitivity rather than angular resolution \citep{quanz2021}, which ultimately facilitates the dection of rocky HZ planets around nearby M dwarfs with \emph{LIFE}  . Especially the high angular resolution of a mid-infrared interferometer enables observations of planets around M dwarfs that are not accessible with a single-aperture near-infrared telescope, further increasing the fraction of detections around (very) late-type stars \citep{kammerer2018}. Consequently, when the yield of Earth-like planets within the HZ of their host star is specifically maximized, an observing sequence that focuses almost entirely ($\sim90\%$) on M dwarfs within 10~pc is obtained \citep{quanz2021}. These findings show that a mission such as \emph{LIFE} could detect $\sim20$ EECs even with smaller 2~m mirrors and demonstrate its eminent scientific capabilities. However, they also raise fundamental questions about the habitability prospects for planets orbiting M dwarfs \citep[e.g.,][]{shields2016} and the performance of a mission such as \emph{LIFE} when focusing entirely on Sun-like stars, as is the case in the \emph{LUVOIR} and \emph{HabEx} reports \citep{luvoir2019,gaudi2020}.
    
    We therefore follow a very conservative approach toward the concept of habitability here and focus solely on the capabilities of a mission such as \emph{LIFE} to detect Earth-like exoplanets in the HZ of Sun-like and FGK-type stars. Therefore, we use the definition of Sun-like stars and the recently published rocky planet occurrence rates specifically inferred for these stars using the Poisson likelihood Bayesian analysis from \citet{bryson2021}. Compared to \citet{quanz2021}, who have published a first yield estimate for the subsample of FGK-type stars, this paper revisits the specific occurrence rate of rocky planets in the HZs of FKG-type and Sun-like stars. The new occurrence rates for rocky planets used in this paper cover the entire optimistic and conservative HZs from \citet{kopparapu2014} for these stars. On the other hand, as illustrated in Section~\ref{sec:habitable_zones}, the planet population used in \citet{quanz2021} misses a significant portion of these HZs, leading to an underestimation of the expected HZ planet yield. Moreover, the new planet occurrence rates from \citet{bryson2021} predict that HZ planets around FGK-type stars are at least as common as those around M dwarfs, resulting in a higher planet yield for the FGK-type star sample than that predicted in \citet{quanz2021}, who used the planet occurrence rates from \citet{kopparapu2018} (i.e., from the ExoPAG SAG13). For a better comparability between the HZ planet yield for FGK-type stars obtained in this paper and the yield published in \citet{quanz2021}, we use the same underlying stellar sample of FGK-type stars as in \citet{quanz2021} in addition to the sample of Sun-like stars as defined in \citet{bryson2021}. Finally, we also go one step further than \citet{quanz2021} and estimate the number of detected HZ planets that might indeed be habitable and harbor liquid surface water using the \texttt{HUNTER} tool from \citet{zsom2015}.
    
    Furthermore, we update our existing planet population synthesis code \texttt{P-pop}\footnote{\url{https://github.com/kammerje/P-pop}} \citep{kammerer2018} to be able to marginalize over the uncertainties in the planet occurrence rate model and reflect these uncertainties in the final yield estimates. Compared to \citet{quanz2021}, we hence obtain more realistic uncertainties and also observe new effects on the observing sequence stemming from the spectral-type dependence of the planet occurrence rates. Future updates of our yield estimate code will include instrumental noise sources in addition to the astrophysical photon noise from stellar leakage, local and exozodiacal dust, and the capability of simulating revisits of promising targets. As outlined in Section~\ref{sec:re-visits}, this will further increase the expected HZ planet yield. We also use our new simulations to study the predicted HZ planet yield as a function of the most important instrumental and astrophysical mission parameters such as the mirror diameter, the total throughput, the spectral coverage, and the exozodiacal dust level. This parameter study enables deriving requirements on the instrument parameters that are driven by the science goals of the mission. Ultimately, this work enables a more direct comparison between thermal emission and reflected-light missions and demonstrates the scientific capabilities of a mission such as \emph{LIFE} under the conservative assumption that life can only exist around Sun-like stars, given that the Earth is still the only habitable planet known to us.
    
    \begin{figure*}
    \centering
    \includegraphics[width=\textwidth]{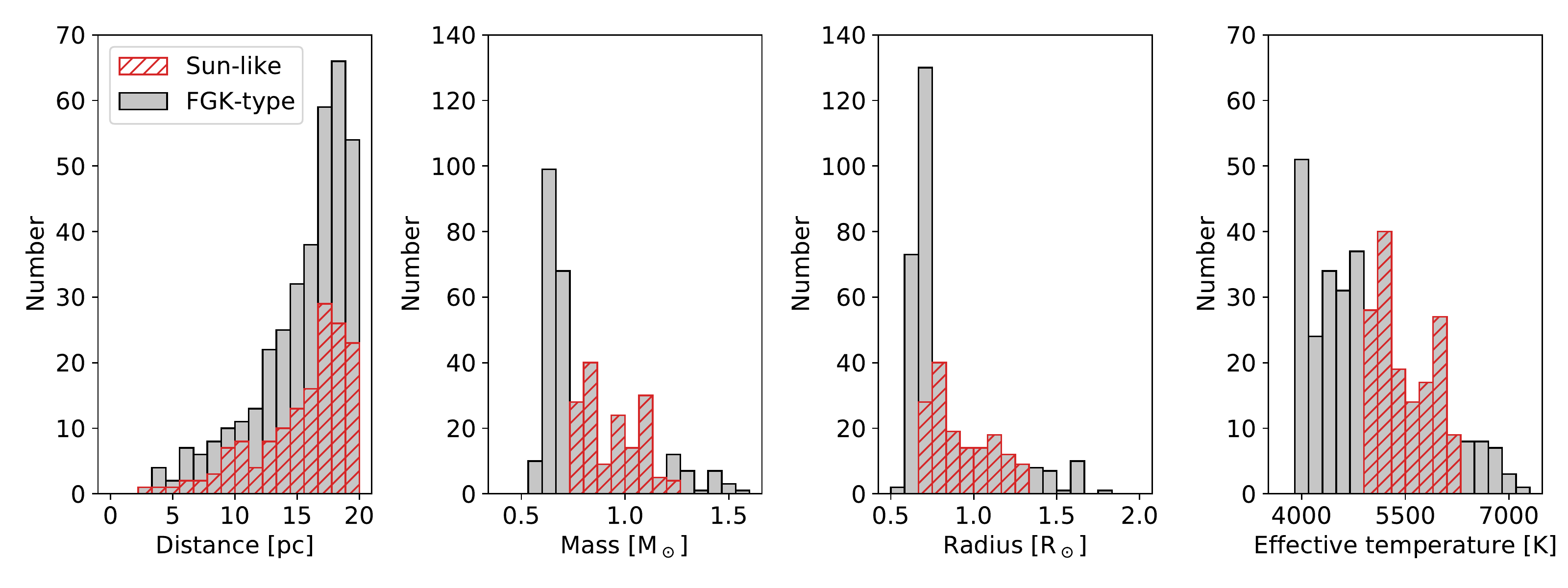}
    \caption{Distance (left), mass (middle left), radius (middle right), and effective temperature (right) distribution of the two stellar samples (Sun-like and FGK-type stars) considered in this work. Sun-like stars are defined as stars with an effective temperature between 4800 and 6300~K. They are a subsample of the FGK-type stars.}
    \label{fig:stellar_sample}
    \end{figure*}
    
    \section{Methods}
    \label{sec:methods}
    
    To estimate the number of Earth-like exoplanets in the HZ of Sun-like and FGK-type stars that can be detected with \emph{LIFE}, we simulated 1'000 synthetic planet populations based on \emph{Kepler} statistics around a sample of nearby (distance $d < 20~\text{pc}$) stars with effective temperatures between 4800--6300~K and 3940--7220~K, respectively. Then, we used \texttt{LIFEsim} \citep{dannert2022} to compute the required integration time for detecting each of the simulated planets assuming photon noise from stellar leakage, local zodiacal, and exozodiacal light. Similar as in \citet{quanz2021}, instrumental noise was not simulated, but we required a conservatively high signal-to-noise ratio (S/N) of 7 integrated over the full wavelength range for detection to account for unconsidered instrumental noise. This corresponds to a 5$\sigma$ detection in a scenario in which photon noise dominates instrumental noise by a factor of 2. For each star in the stellar sample, we then computed the amount of time that it takes to detect each of the planets simulated around it. The integration time was then assigned to the different stars in the order of decreasing efficiency, where the most efficient observation is the one that detects the largest number of planets per integration time. By focusing on only a subset of the simulated planets, this enables constructing a single-visit observing sequence that maximizes the expected yield of Earth-like exoplanets in the HZ of their host stars, similar to the altruistic yield optimization routine in \citet{stark2014}. By averaging over 1'000 synthetic planet populations, we marginalized over different numbers and properties of simulated planets, orbital realizations, and exozodiacal dust levels. Re-visits to make use of the orbital motion of the planets and detect those that might have been too close to their host star to be resolved in the first visit were not simulated, but could further increase the estimated planet yield beyond the predictions in this work (see Section~\ref{sec:re-visits}). The individual steps of our simulations are outlined in the following sections.
    
    \subsection{Stellar sample}
    \label{sec:stellar_sample}
    
    We used the same stellar sample as \citet{quanz2021} as host stars for the synthetic planet populations. This sample consists of single and wide binary main-sequence stars within 20~pc from the Sun. The main focus of this work is the sample of Sun-like stars, defined as stars with an effective temperature between 4800 and 6300~K (roughly corresponding to main-sequence stars with spectral types between K3V and F7V) according to \citet{bryson2021}, but for better comparison with the yield estimates presented in \citet{quanz2021}, we also considered the sample of all FGK-type stars with effective temperatures between 3940 and 7220~K (of which Sun-like stars are a subsample). These samples consist of 154 Sun-like and 358 FGK-type stars, of which 141 and 316 are single stars, and the remaining 13 and 42 are wide binary stars, respectively. The distance, mass, radius, and effective temperature distribution of both samples is shown in Figure~\ref{fig:stellar_sample}. We note that $\alpha$~Cen~A and~B are not included in these samples.
    
    \subsection{Planet population}
    \label{sec:planet_population}
    
    \begin{figure}
        \centering
        \includegraphics[width=\columnwidth]{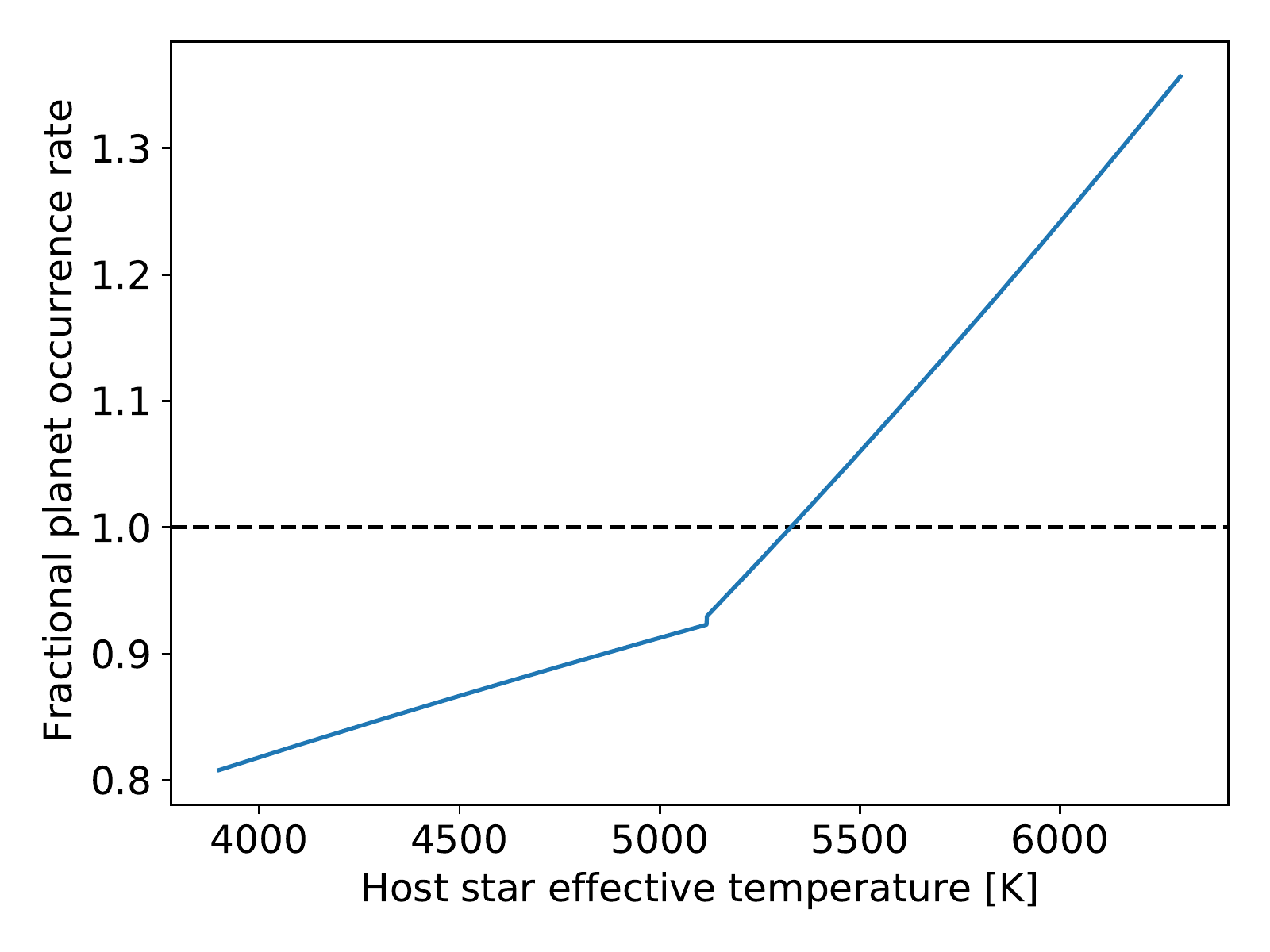}
        \caption{Fractional planet occurrence rate as a function of the host star effective temperature for the median model parameters of the model 1 hab2 stars low bound with uncertainty (hab2min) case from Table~1 of \citet{bryson2021}. We note that the average value of this function over $T_\text{eff}$ from 3900 to 6300~K is one.}
        \label{fig:n0_teff_dependence}
    \end{figure}
    
    To generate the synthetic exoplanet populations, we used the publicly available planet population synthesis tool \texttt{P-pop} \citep{kammerer2018} with recent occurrence rate estimates for rocky planets from \citet{bryson2021}, which are based on the \emph{Kepler} DR25 planet candidate catalog \citep{thompson2018} and \emph{Gaia}-based stellar properties \citep{gaia2018}. Here, we consider two cases: the model 1 hab2 stars low bound (hab2min) and the model 1 hab2 stars high bound (hab2max), both with uncertainty, from Table~1 of \citet{bryson2021}. The low and high bounds refer to different extrapolation scenarios for orbital periods beyond 500~days, for which data supporting completeness characterization are not available in \emph{Kepler} DR25 \citep{thompson2018}. A conservative extrapolation is given by the low-bound scenario, in which completeness for orbital periods beyond 500~days is set to the completeness at 500~days, whereas an optimistic extrapolation is given by the high-bound scenario, in which completeness beyond 500~days is set to zero. Moreover, we decided to consider the hab2 stars sample because it is based on a larger number of planet candidates and on a wider range of stellar effective temperatures. This sample contains 68'885 stars with effective temperatures ranging from 3900 to 6300~K. \citet{bryson2021} removed poorly characterized, binary, evolved stars, and stars whose observations were not well suited for long-period transit searches from the sample, as outlined in Section~3.1 of their paper. The majority of the 68'885 remaining stars in the sample has radii smaller than $1.35~\text{R}_\odot$ , as recommended by \citet{burke2017}, but instead of a radius cut, \citet{bryson2021} opted for a physically motivated selection based on the evolutionary flag, leaving 6.8\% of stars with radii larger than $1.35~\text{R}_\odot$ in the sample. They did not apply a transit duration cut at 15~hours either, as is done in the \emph{Kepler} pipeline, because a significant fraction of the stars in the hab2 sample has transit durations of more than 15~hours at the outer edge of the optimistic HZ. However, they noted that longer transit durations either have a small impact on the detection completeness or decrease it so that their impact falls within the considered low- and high-bound cases. We focused on model 1 because it has more parameters to trace the dependence of the planet occurrence rate on the stellar effective temperature. This dependence is illustrated for the hab2min case in Figure~\ref{fig:n0_teff_dependence}. While \citet{bryson2021} mentioned that the dependence of the planet occurrence rate on the spectral type is not statistically significant and thus only a tentative detection, we are aware that including this spectral-type dependence in our simulations may introduce biases in our results. However, because the planet occurrence rate is assumed to be increasing with increasing stellar effective temperature while the number of stars in our stellar sample is decreasing with increasing stellar effective temperature, our simulations should still be on the conservative side. The two considered cases also account for the uncertainties in the planet candidate radius, instellation flux (i.e., host star flux incident on the planet), and host star effective temperature and are described by the planet distribution model,
    \begin{equation}
        \label{eqn:model}
        \frac{\partial^2N(R_\text{p},F_\text{p},T)}{\partial R_\text{p}\partial F_\text{p}} = N_0CR_\text{p}^\alpha F_\text{p}^\beta T^\gamma g(T),
    \end{equation}
    where $N$ is the average number of planets per star, $R_\text{p}$ is the planet radius, $F_\text{p}$ is the planet instellation flux, $T = T_\text{eff}/\text{T}_\odot$ with $T_\text{eff}$ the host star effective temperature and $\text{T}_\odot = 5778~\text{K}$ the effective temperature of the Sun, and
    \begin{align}
        g(T_\text{eff}) = \begin{cases} 10^{-11.84}T_\text{eff}^{3.16} \qquad \text{if }T_\text{eff} \leq 5117~\text{K}, \\ 10^{-16.77}T_\text{eff}^{4.49} \qquad \text{otherwise}. \end{cases}
    \end{align}
    We note that $N_0$, $\alpha$, $\beta$, and $\gamma$ are the model parameters from Table~1 of \citet{bryson2021}, and $C$ was chosen so that
    \begin{equation}
        \int_{0.5~\text{R}_\oplus}^{2.5~\text{R}_\oplus}\int_{0.2~\text{F}_\oplus}^{2.2~\text{F}_\oplus}\frac{\partial^2N(R_\text{p},F_\text{p},T)}{\partial R_\text{p}\partial F_\text{p}}\text{d}R_\text{p}\text{d}F_\text{p} = N_0
    \end{equation}
    if averaged over $T_\text{eff}$ from 3900 to 6300~K, where $\text{R}_\oplus$ is the radius of the Earth and $\text{F}_\oplus$ is the instellation flux received by the Earth. This means that the planet occurrence rate integrated over $R_\text{p}$ from 0.5 to $2.5~\text{R}_\oplus$ and $F_\text{p}$ from 0.2 to $2.2~\text{F}_\oplus$ for a given host star with effective temperature $T_\text{eff}$ is given by
    \begin{equation}
        N_{T_\text{eff}} = N_0C_T\left(\frac{T_\text{eff}}{\text{T}_\odot}\right)^\gamma g(T_\text{eff}),
    \end{equation}
    where
    \begin{equation}
        C_T = \left(\frac{\int_{3900~\text{K}}^{6300~\text{K}}T^\gamma g(T)\text{d}T}{6300~\text{K}-3900~\text{K}}\right)^{-1}
    \end{equation}
    is the inverse of the average of the host star effective temperature-dependent part of Equation~\ref{eqn:model}. Hence, for each star in our sample, we drew a random number of planets from a Poisson distribution with mean $N_{T_\text{eff}}$ and randomly distribute the planet radius and instellation flux according to Equation~\ref{eqn:model} by feeding uniformly distributed random numbers between 0 and 1 into the normalized inverse cumulative distribution functions of the planet radius- and instellation flux-dependent parts of Equation~\ref{eqn:model}. Then, we repeated this step 1'000 times to simulate 1'000 synthetic planet populations. We marginalized over the uncertainties in the model parameters $N_0$, $\alpha$, $\beta$, and $\gamma$ by drawing a random set of parameters for each of the 1'000 synthetic planet populations from the Markov-Chain Monte Carlo (MCMC) posteriors of \citet{bryson2021} (obtained via private communication). This approach enables reflecting the uncertainties in the underlying planet distribution in our final yield estimates. The uncertainties introduced by drawing from a Poisson distribution are given by $\sqrt{N_{T_\text{eff}}/1'000}$ and are negligible compared to those in the underlying planet distribution.
    
    Based on results from \citet{kraus2016}, we scaled the planet occurrence rate around close binary stars with separations $<50~\text{au}$ to 30\% of the nominal rate around single stars. Unlike in \citet{quanz2021}, we did not apply a multiplanet system stability criterion to ensure that a randomly drawn planetary system is stable. The reason for this choice is that if a high value ($\gtrapprox5$) is drawn for $N_0$ from the MCMC posterior of \citet{bryson2021}, it takes too long to randomly draw a stable multiplanet system (because a very dense and regular orbital period and planet mass spacing would be required). Moreover, redrawing multiplanet systems until they happen to be stable leads to deviations of the simulated from the desired planet radius and instellation flux distributions and biases these distributions toward smaller planets and more widely spaced orbits. Part of this problem might also be the tail of high planet occurrence rates in the MCMC posterior distribution itself, which could predict too many planets within a given orbital period range for a system to be stable. This might be the case because \citet{bryson2021} did not perform a system stability analysis when they derived their planet occurrence rates. While we end up with a significant fraction of unstable multiplanet systems, it is important to note that the stability of the systems is completely irrelevant for our yield estimates because \texttt{LIFEsim} treats each individual planet as a separate observation and we did not consider revisits, on which the orbital motion of multiplanet systems would have an impact. Instead, it is only relevant that we simulate a planet population whose combined planet radius and instellation flux distributions match those predicted by \citet{bryson2021}.
    
    After the synthetic planets were generated, we assigned them randomly oriented (but coplanar within a system) circular orbits and uniformly distributed Bond and geometric albedos between 0 and 0.8 (mean 0.4) and 0 and 0.1 (mean 0.05), respectively. We note that we assumed the wavelength-dependent geometric albedo to be constant over the entire \emph{LIFE} wavelength regime, but because we considered wavelengths longer than $3~\text{\textmu m}$, the reflected host star light (and therefore the geometric albedo) is negligible compared to the thermal emission, at least for the small and mature planets that are of interest here. All planets were considered to emit like blackbodies with a radius $R_\text{p}$ and an effective temperature $T_\text{p}$ that depends on the luminosity of their host star, their Bond albedo, and their orbital separation, similar as in \citet{kammerer2018}. Moreover, to account for the photon noise from exozodiacal light, we assigned each planetary system a randomly drawn zodi level according to either the nominal (median of $\sim3$ zodi) or the pessimistic (median of $\sim12$ zodi) exozodi level distributions from the HOSTS survey \citep{ertel2018,ertel2020}.
    
    \subsection{Habitable zones}
    \label{sec:habitable_zones}
    
    \begin{figure}
        \centering
        \includegraphics[width=\columnwidth]{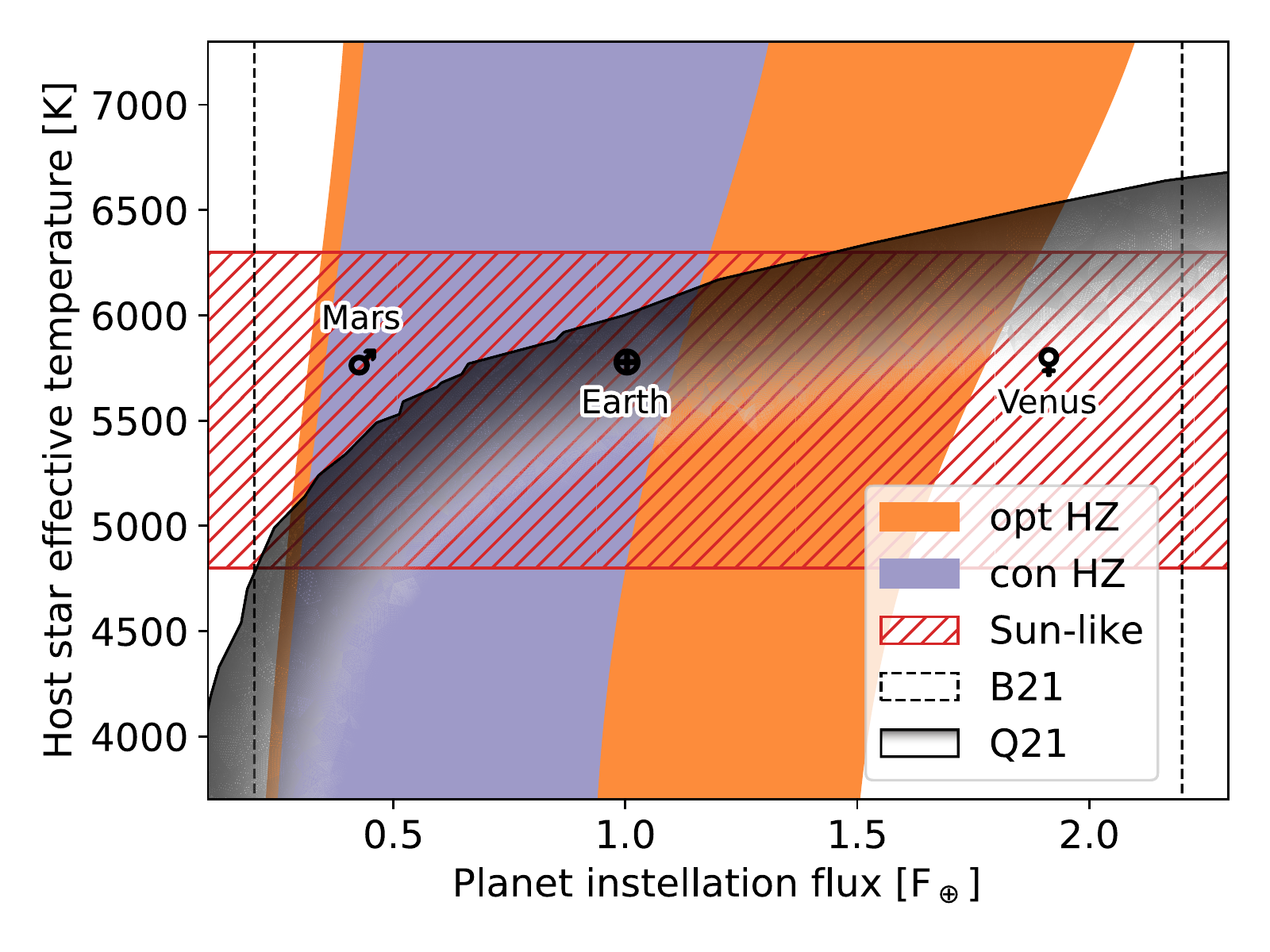}
        \caption{HZs considered in this work in the planet instellation flux vs. host star effective temperature plane. The region with dashed black edge shows the parameter space in which planets are simulated in this work using the \citet{bryson2021} planet occurrence rates. The region with solid black and shaded edge shows the parameter space in which planets are simulated in \citet{quanz2021}. The black symbols highlight the positions of the Earth, Mars, and Venus in this figure, and the dashed red region shows the parameters space of Sun-like stars according to \citet{bryson2021}.}
        \label{fig:habitable_zones}
    \end{figure}
    
    \begin{table*}
    \caption{Definition of the three types of exoplanets with their occurrence rate considered in this work: rocky planets within the opt HZ, rocky planets within the con HZ, and exo-Earth candidates (ECCs). Shown are the average number of planets per Sun-like ($\eta$) and FGK-type ($\zeta$) star in the hab2min and hab2max planet populations.}
    \label{tab:habitable_zones}
    \centering
    \begin{tabular}{ccccccc}
    \hline\hline
    Name & $R_\text{p}$ [$\text{R}_\oplus$] & $F_\text{p}$ [$\text{F}_\oplus$] & $\eta_\text{hab2min}$ & $\eta_\text{hab2max}$ & $\zeta_\text{hab2min}$ & $\zeta_\text{hab2max}$\\
    \hline
    opt HZ & 0.5--1.5 & 0.320--1.776$^{(a)}$ & $0.56^{+0.68}_{-0.32}$ & $0.93^{+1.15}_{-0.56}$ & $0.53^{+0.58}_{-0.28}$ & $0.80^{+0.97}_{-0.45}$\\
    con HZ & 0.5--1.5 & 0.356--1.107$^{(b)}$ & $0.35^{+0.46}_{-0.20}$ & $0.62^{+0.86}_{-0.39}$ & $0.34^{+0.39}_{-0.19}$ & $0.54^{+0.72}_{-0.31}$\\
    EEC & 0.8$^{(c)}$--1.4 & 0.356--1.107$^{(b)}$ & $0.19^{+0.19}_{-0.10}$ & $0.34^{+0.38}_{-0.19}$ & $0.15^{+0.14}_{-0.08}$ & $0.26^{+0.32}_{-0.15}$\\
    \hline
    \multicolumn{7}{l}{\textbf{Notes.} \parbox[t]{11 cm}{$R_\text{p}$ denotes the planet radius and $F_\text{p}$ denotes the planet instellation flux. Occurrence rates are averaged over our input catalog for the given range of spectral types. Because the stellar flux range and the underlying occurrence rate model is depends on spectral type, the number of objects falling within this range, and hence the occurrence rate, varies with spectral type as well. (a) Dependent on spectral type according to the ``Early Mars'' and ``Recent Venus'' scenarios in \citet{kopparapu2014}. (b) Dependent on spectral type according to the ``Maximum Greenhouse'' and ``Runaway Greenhouse'' scenarios in \citet{kopparapu2014}; only the values for $1~\text{M}_\oplus$ planets are adopted here. (c) For EECs, the lower limit of the radius range depends on the separation from the host star; $R_\text{p}^\text{min} = 0.8a_\text{p}^{-0.5}~\text{R}_\oplus$, where $a_\text{p}$ is the planet semi-major axis in units of au \citep[see][]{stark2019}.}}
    \end{tabular}
    \end{table*}
    
    The HZ is commonly defined as the region around a star in which liquid surface water can exist on a rocky planet that has an atmospheric composition and climate stability feedback similar to our own Earth \citep{huang1959,hart1978,kasting1993,kasting2003}. Previous works \citep{luvoir2019,gaudi2020,quanz2021} considered an optimistic HZ (opt HZ) based on the so-called early Mars and recent Venus scenarios and a conservative HZ (con HZ) based on the so-called maximum greenhouse and runaway greenhouse scenarios from \citet{kopparapu2014}. These scenarios define two different planet instellation flux ranges for the HZ. We note that both opt HZ limits take the luminosity evolution of the Sun into account, which was fainter during the epochs when Venus and Mars provided habitable conditions. For the present-day solar luminosity, these insolation limits correspond to separations of 0.75 and 1.77~au, respectively, excluding Venus from the opt HZ, but including Mars. The planet radius range of the HZ is usually defined as $0.5~\text{R}_\oplus \leq R_\text{p} \leq 1.5~\text{R}_\oplus$. A third case called exo-Earth candidates (EECs) with a planet instellation flux range similar to the con HZ but with a tighter planet radius range has also been considered, however. For this case of EECs, the lower limit of the radius range is given by $R_\text{p}^\text{min} = 0.8a_\text{p}^{-0.5}~\text{R}_\oplus$, where $a_\text{p}$ is the planet semi-major axis in units of au \citep[see][]{stark2019}, so that planets closer to the star are required to have a larger radius in order to be able to hold on to an atmosphere and be potentially habitable. For better comparability with other works, we adopt the same three definitions of the HZ. They are also described in more detail in Table~\ref{tab:habitable_zones}.
    
    Proper interpretation of the final yield estimates requires understanding of detection biases introduced by the range of simulated planets that only partially overlap with the HZs. The planet population from \citet{bryson2021} simulates planet radii between 0.5 and $2.5~\text{R}_\oplus$ and planet instellation fluxes between 0.2 and $2.2~\text{F}_\oplus$ . It therefore covers the entire con HZ, opt HZ, and the parameter space defined as exo-Earth candidates (see Figure~\ref{fig:habitable_zones}). However, the planet population from \citet{kopparapu2018} that was used in \citet{quanz2021} only simulates planet orbital periods between 0.5 and 500~days. When this is converted into planet instellation flux for the stars in the FGK-type sample, it does not cover a significant part of the HZ around earlier-type stars (see the opt HZ and con HZ regions that do not overlap with the Q21 region in Figure~\ref{fig:habitable_zones}). This needs to be kept in mind when yield estimates from this work are compared to those in \citet{quanz2021}. 
    \subsection{Mission parameters}
    \label{sec:mission_parameters}
    
    \begin{table}
    \caption{Mission parameters adopted for the space-based nulling interferometer.}
    \label{tab:mission_parameters}
    \centering
    \begin{tabular}{llll}
    \hline\hline
    \multicolumn{2}{l}{Parameter} & Value & Unit\\
    \hline
    \multicolumn{2}{l}{Number of collector spacecraft} & 4 & --\\
    \multirow{3}{*}{Mirror diameter} & pessimistic & 1.0 & m\\
    & reference & 2.0 & m\\
    & optimistic & 3.5 & m\\
    \multirow{3}{*}{Wavelength coverage} & pessimistic & 6.0--17.0 & $\text{\textmu m}$\\
    & reference & 4.0--18.5 & $\text{\textmu m}$\\
    & optimistic & 3.0--20.0 & $\text{\textmu m}$\\
    \multicolumn{2}{l}{Spectral resolution} & 20 & --\\
    \multicolumn{2}{l}{Min. nulling baseline length} & 10 & m\\
    \multicolumn{2}{l}{Max. nulling baseline length} & 100 & m\\
    \multicolumn{2}{l}{Ratio imaging$/$nulling baseline} & 6 & --\\
    \multicolumn{2}{l}{Quantum efficiency} & 0.7 & --\\
    \multicolumn{2}{l}{Total throughput} & 0.05 & --\\
    \multicolumn{2}{l}{$\text{S/N}_\text{target}$} & 7 & --\\
    \multicolumn{2}{l}{Duration of search phase} & 2.5 & yr\\
    \multicolumn{2}{l}{Slew time} & 10 & h\\
    \multicolumn{2}{l}{Rel. observing overheads} & 0.2 & --\\
    \hline
    \end{tabular}
    \end{table}
    
    The mission parameters adopted for the space-based nulling interferometer are the same as in \citet{quanz2021} and can be found in Table~\ref{tab:mission_parameters}. The array architecture is based on \citet{cockell2009} and consists of four free-flying collector spacecraft that send their light to a fifth beam-combiner spacecraft. For the collector spacecraft mirror diameter and the wavelength coverage, we considered three different scenarios (pessimistic, reference, and optimistic). In the remainder of this work, these three scenarios are only referred to with their corresponding mirror diameters (1, 2, and 3.5~m), and the corresponding wavelength regimes are adopted implicitly. We note that in addition to the mirror diameter, the total throughput (here assumed to be 5\%) also has a strong impact on the final yield estimates. While 5\% is a rather conservative choice \citep[the Large Binocular Telescope Interferometer (LBTI) currently achieves a total throughput of $\sim11\%$ at $11~\text{\textmu m;}$][]{quanz2021}, the dependence of the predicted planet yield on the total throughput and also the mirror diameter are more closely investigated in Section~\ref{sec:mission_parameter_study}. Because \texttt{LIFEsim} does not yet account for instrumental noise, we require a conservatively high signal-to-noise ratio (S/N) of 7 for detection. The duration of the search phase was set to 2.5~years, and the remaining 2.5~years of a nominal 5-year mission could be spent on in-depth spectral characterization of the most promising targets. The slew time was set to a constant value of 10~hours to simulate a penalty for switching targets, and the relative observing overheads were assumed to be 20\%. While we used \texttt{LIFEsim} to compute an individually optimized observing sequence that maximizes the yield of HZ planets for each of the considered scenarios, we visited each system only once. We note that to determine the orbits of any detected planets and to reject false positives, multiple visits will eventually be required in advance of an in-depth characterization phase.
    
    \section{Results}
    \label{sec:results}
    
    Before we studied the yield estimates, we performed a quick sanity check by comparing the average number of planets per star in our synthetic planet populations to those reported in \citet{bryson2021}. For our two synthetic planet populations (hab2min and hab2max) around Sun-like stars, we find an average number of planets per star with $R_\text{p} \in [0.5,1.5]~\text{R}_\oplus$ and $F_\text{p} \in [0.2,2.2]~\text{F}_\oplus$ of $1.13^{+0.95}_{-0.49}$ and $1.87^{+1.97}_{-0.93}$, respectively, of which $0.35^{+0.46}_{-0.20}$ ($0.37^{+0.48}_{-0.21}$) and $0.62^{+0.86}_{-0.39}$ ($0.60^{+0.90}_{-0.36}$) reside within the con HZ and $0.56^{+0.68}_{-0.32}$ ($0.58^{+0.73}_{-0.33}$) and $0.93^{+1.15}_{-0.56}$ ($0.88^{+1.28}_{-0.51}$) reside within the opt HZ (in parentheses are the numbers from \citet{bryson2021} for comparison). The small differences between our and their numbers arise because our stellar sample is different (the planet occurrence rate depends on the effective temperature of the host star) and because of the statistical nature of our simulations.
    
    \subsection{Planet yield of Sun-like stars }
    \label{sec:sun-like_stars_planet_yield}
    
    \begin{figure*}
        \centering
        \includegraphics[width=\textwidth]{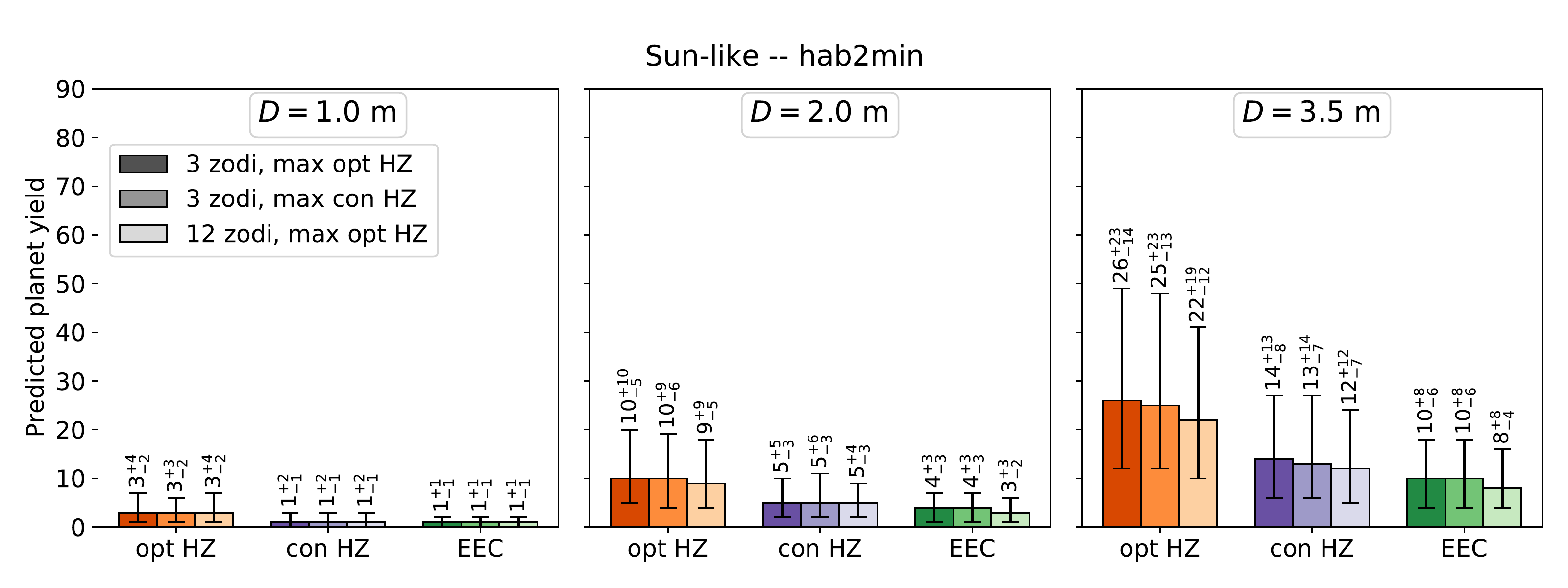}
        \includegraphics[width=\textwidth]{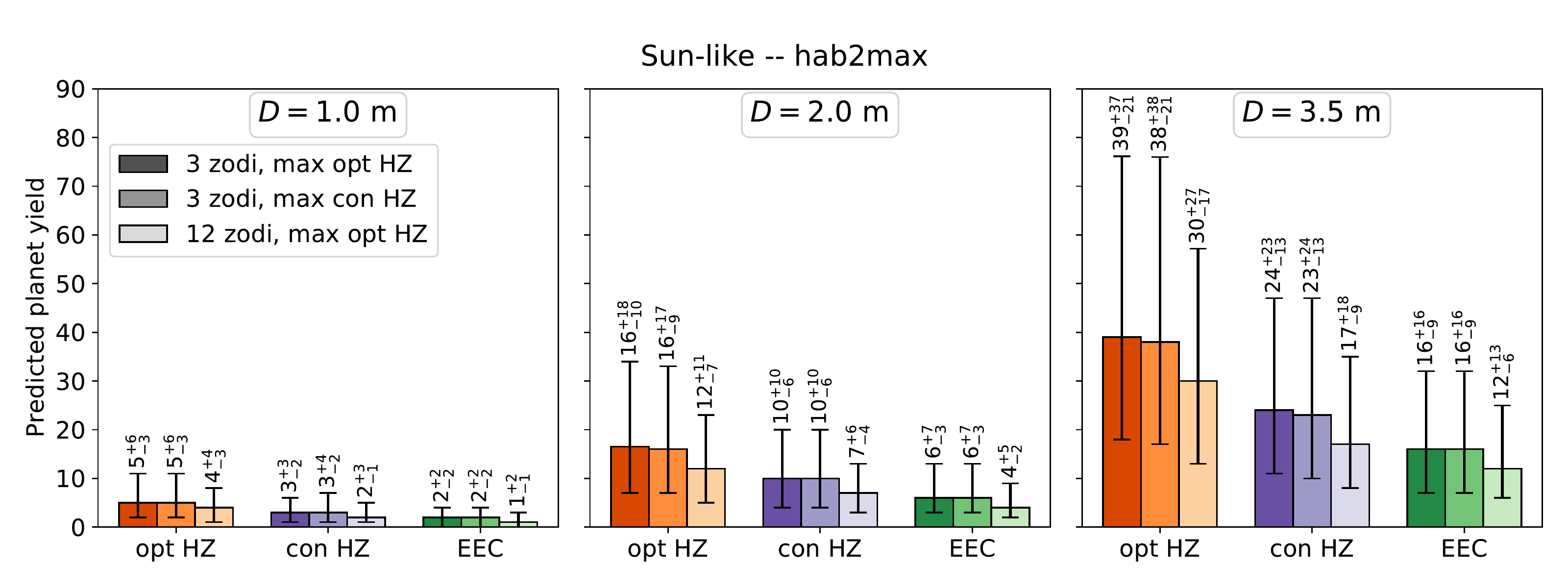}
        \caption{Predicted planet yield for the sample of Sun-like stars. Each of the six panels shows the predicted yield for the three different HZ definitions in different colors (see Section~\ref{sec:habitable_zones}) as well as for three different scenarios in different shadings: (1) median exozodiacal dust level of 3~zodi and interferometric baseline length optimization for the center of the opt HZ, (2) median exozodiacal dust level of 3~zodi and interferometric baseline length optimization for the center of the con HZ, and (3) median exozodiacal dust level of 12~zodi and interferometric baseline length optimization for the center of the opt HZ. The top row is based on the hab2min and the bottom row is based on the hab2max planet populations (see Section~\ref{sec:planet_population}), while the left, middle, and right columns show the predicted yields for the 1, 2, and 3.5~m mirrors, respectively.}
        \label{fig:yield_sun}
    \end{figure*}
    
    One of the scientific goals that was formulated for the \emph{LIFE} mission is that at least 30 (requirement) / 50 (goal) extrasolar planets with radii between 0.5 and 1.5 Earth radii are to be investiated. They should receive between 0.35 and 1.7 times the instellation flux of the Earth in order to assess their habitability and search their atmospheres for biosignatures \citep{quanz2021_whitepaper}. We call these planets the potentially habitable planets here. As outlined in Section~\ref{sec:habitable_zones}, we consider three different definitions of a potentially habitable planet: (1) a planet within the opt HZ, (2) a planet within the con HZ, and (3) an exo-Earth candidate.
    
    Table~\ref{tab:planet_yield} shows the total predicted yield including uncertainties for the three different potentially habitable planet types mentioned above, as well as for all simulated planets (i.e., $0.5~\text{R}_\oplus \leq R_\text{p} \leq 2.5~\text{R}_\oplus$ and $0.2~\text{F}_\oplus \leq F_\text{p} \leq 2.2~\text{F}_\oplus$) for the sample of Sun-like stars ($4800~\text{K} \leq T_\text{eff} \leq 6300~\text{K}$). We note that for the remainder of this paper, ``total predicted yield'' means the median number of detectable planets over the 1'000 simulated synthetic universes, and the uncertainties reflect the 16th and 84th percentiles of this distribution. In total, we present 24 different scenarios that are based on two different extrapolation models for the planet occurrence rate  (hab2min and hab2max), two different exozodiacal dust level distributions (with a median of 3 and 12~zodi), two different optimization schemes for the interferometric baseline length (center of the opt HZ and center of the con HZ), and three different mirror diameters (1, 2, and 3.5~m; we note that the wavelength regime is adapted implicitly, see Section~\ref{sec:mission_parameters}). In all cases, the total instrument throughput was set to 5\%. We note that the modulation map of a four-telescope nulling interferometer is a pattern with many transmission minima and maxima \citep[e.g.,][]{dannert2022}. If the telescopes are free flying, their distances (i.e., baselines) can be optimized so that the first transmission maximum occurs at the center of the respective HZ for a face-on system. Figure~\ref{fig:yield_sun} compares the predicted planet yield for several of these scenarios. The main finding is that if \emph{LIFE} were to only target Sun-like stars, 3.5~m mirrors are required to detect 30 potentially habitable planets under the assumptions laid out in Table~\ref{tab:mission_parameters}, even for the opt HZ. However, the 1$\sigma$ upper bounds case would even fulfill the mission goal of detecting 50 potentially habitable planets with 3.5~m mirrors or the mission requirement of detecting 30 potentially habitable planets with 2~m mirrors. Nevertheless, these findings should be taken with caution because allowing for revisits would further increase the predicted planet yield by a significant fraction (see Section~\ref{sec:re-visits}). Furthermore, the average yield with 2~m mirrors is approximately the same as the case with 1$\sigma$ lower bounds with 3.5~m mirrors and would probably be at the lower limit of what could be considered a statistically meaningful sample. This is because with a sample of $\text{about ten}$ potentially habitable planets, the habitability fraction of the Solar System (which is defined as the fraction of HZ planets that are actually habitable, and is 50\%) could be rejected with a significance of $\sim3\sigma$ in case of a null result \citep{quanz2021_whitepaper}. Finally, while several potentially habitable planets could be detected with 1m mirrors, a statistically meaningful assessment of the habitability fraction would likely be impossible.
    
    Our parameter study also reveals the impact of the interferometric baseline length optimization scheme and the median exozodiacal dust level on the predicted planet yield. Figure~\ref{fig:yield_sun} shows that while the former has virtually no impact on the planet yield, the latter has a small but noticeable impact that increases with increasing mirror size. This is because with larger mirrors, exoplanetary systems at greater distances can be observed, whose the exozodiacal light is brighter than that of the planets. The uncertainties on the exozodiacal dust levels are still large, therefore it is encouraging to see that \emph{LIFE} would only be weakly affected by a rather pessimistic distribution of this inevitable source of noise. We note that \texttt{LIFEsim} currently assumes a perfect background (and therefore exozodiacal light) subtraction down to the photon noise limit. While this might be an overly optimistic assumption for reflected-light missions, where exozodiacal dust disks will appear highly asymmetric due to the different forward and backward scattering cross-sections of the dust \citep[e.g.,][]{stark2014_hg}, subtraction of the exozodiacal dust disk should be more easy in the thermal infrared, where the near and far side of the disk will appear more symmetric. Nevertheless, potential substructure or disk offsets would also lead to additional disk flux leaking through the interferometric modulation map, although \citet{defrere2010} found that this should not be an issue for disks with exozodi levels smaller than 50~zodi and offsets smaller than 0.5~au.
    
    \subsection{Planet yield of FGK-type stars}
    \label{sec:fgk-type_stars_planet_yield}
    
    \begin{figure*}
        \centering
        \includegraphics[width=\textwidth]{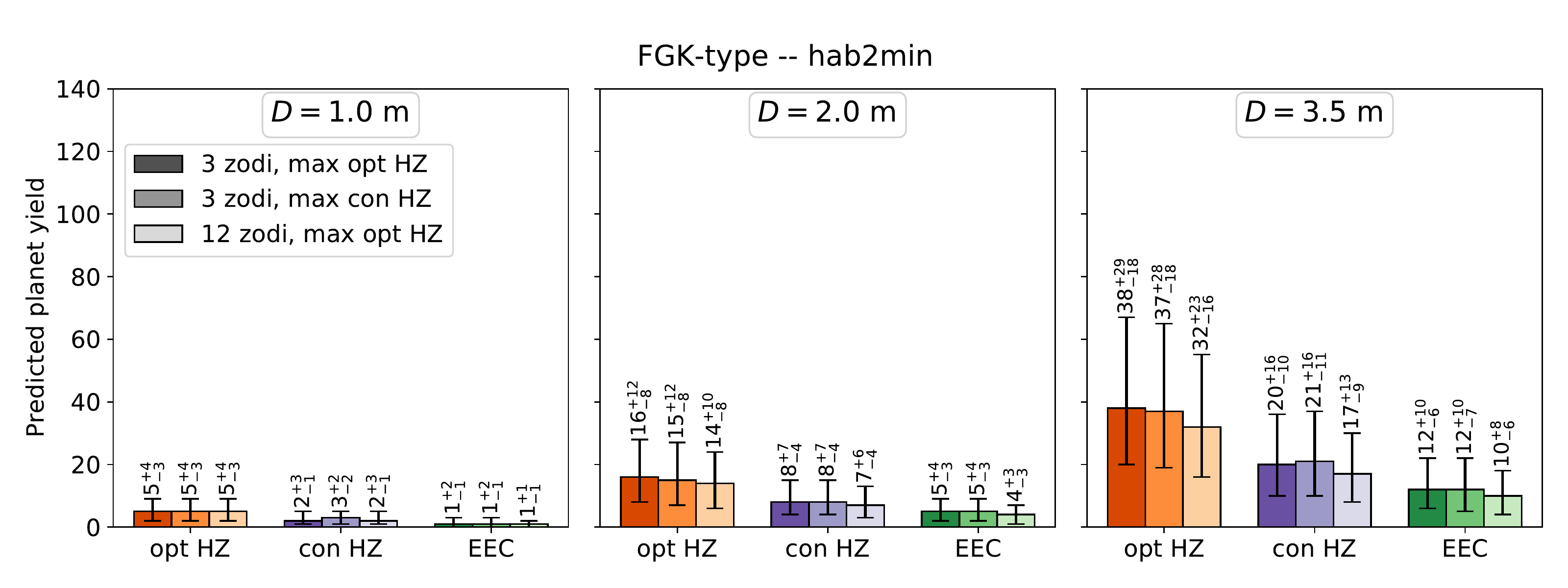}
        \includegraphics[width=\textwidth]{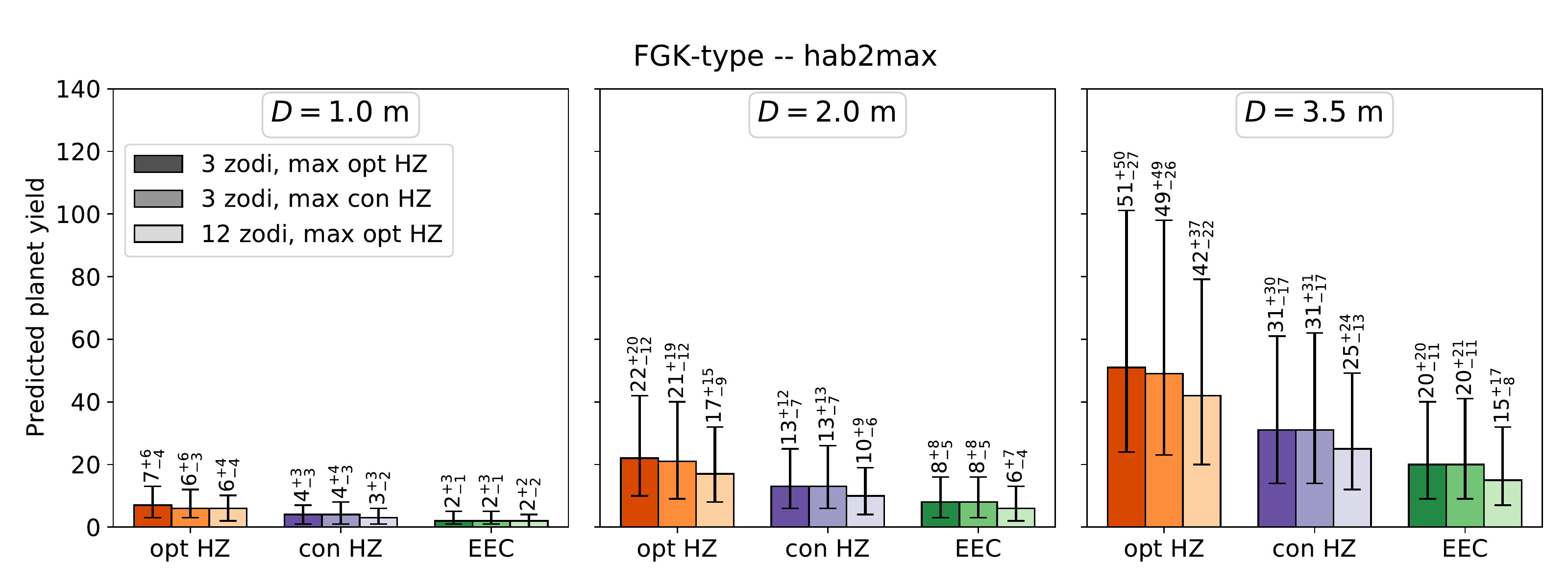}
        \caption{Same as Figure~\ref{fig:yield_sun}, but for the sample of FGK-type stars.}
        \label{fig:yield_fgk}
    \end{figure*}
    
    Whether only Sun-like stars as defined in \citet{bryson2021} and considered in the previous Section~\ref{sec:sun-like_stars_planet_yield} or more broadly all FGK-type stars should be considered in a search for potentially habitable planets remains subject to debate, similar to whether M-type stars should also be investigated \citep[e.g.,][]{scalo2007,shields2016}. \citet{quanz2021} have presented yield predictions for the subset of FGK-type stars within 20~pc from the Sun. For a direct comparison, we also present yield predictions for the same sample of FGK-type stars here. As noted above, the sample of Sun-like stars is a subset of the sample of FGK-type stars. While the underlying planet population in \citet{quanz2021} is based on the results of the NASA ExoPAG SAG13 \citep{kopparapu2018} and misses part of the HZ around early-type stars, the results presented here are based on the planet population from \citet{bryson2021}, which covers the entire con HZ and opt HZ (see Figure~\ref{fig:habitable_zones}). Hence, the underlying planet occurrence rates in the opt HZ and con HZ are significantly higher here (see Table~\ref{tab:planet_occurrence_rates}).
    
    \begin{table}
    \caption{Comparison of the underlying planet occurrence rates for a Sun-like star in \citet{quanz2021} and the average of the FGK-type sample in this work.}
    \label{tab:planet_occurrence_rates}
    \centering
    \begin{tabular}{cccc}
    \hline\hline
    Name & $\eta$\citep{quanz2021} & $\eta_\text{hab2min}$ & $\eta_\text{hab2max}$\\
    \hline
    opt HZ & 0.32 & 0.53 & 0.80\\
    con HZ & 0.17 & 0.34 & 0.54\\
    \hline
    \end{tabular}
    \end{table}
    
    Figure~\ref{fig:yield_fgk} shows the predicted planet yield for the sample of FGK-type stars, similar to Figure~\ref{fig:yield_sun} for the sample of Sun-like stars. While the predicted yield is similar to that in \citet{quanz2021} for EECs and slightly higher for planets within the opt HZ with the hab2min planet population, there is a significant increase in predicted yield among all potentially habitable planet types with the hab2max planet population. This increase in predicted yield is roughly consistent with the increase in the underlying planet occurrence rates between \citet{quanz2021} and this work, that is, $\sim100\%$ with the hab2min and $\sim200\%$ with the hab2max planet population, respectively. The numbers shown in Table~\ref{tab:planet_occurrence_rates} slightly overestimate the increase in predicted yield because the underlying planet occurrence rates in \citet{quanz2021} are given for a Sun-like star, while the sample of FGK-type stars contains significantly more later-type stars (see Figure~\ref{fig:stellar_sample}). For these later-type stars, the underlying planet occurrence rates in \citet{quanz2021} cover a larger fraction of the con HZ and opt HZ, so that the difference to the planet occurrence rates used in this work shrinks. However, the most important finding is that if \emph{LIFE} would target all FGK-type stars within 20~pc (and not just the Sun-like stars), then the mission would approach its goal of detecting 50 potentially habitable planets with 3.5~m mirrors even in an average scenario, and would still be able to reach its requirement of detecting 30 potentially habitable planets with smaller mirrors somewhere between 2 and 3.5~m in size. We note that we study the dependence of the predicted planet yield on the mirror size
in more detail  in Section~\ref{sec:mission_parameter_study}.
    
    \subsection{Optimization of the observing sequence}
    \label{sec:observing_sequence_optimization}
    
    \begin{figure*}
        \centering
        \includegraphics[width=\textwidth]{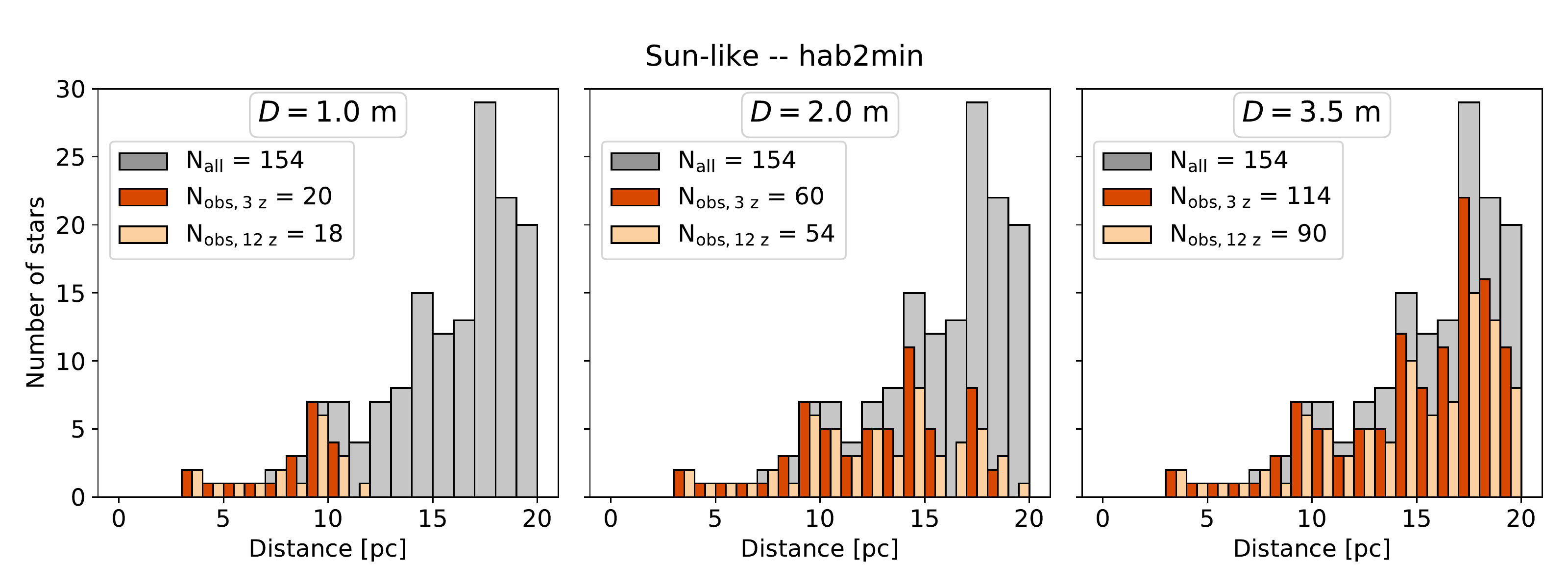}
        \includegraphics[width=\textwidth]{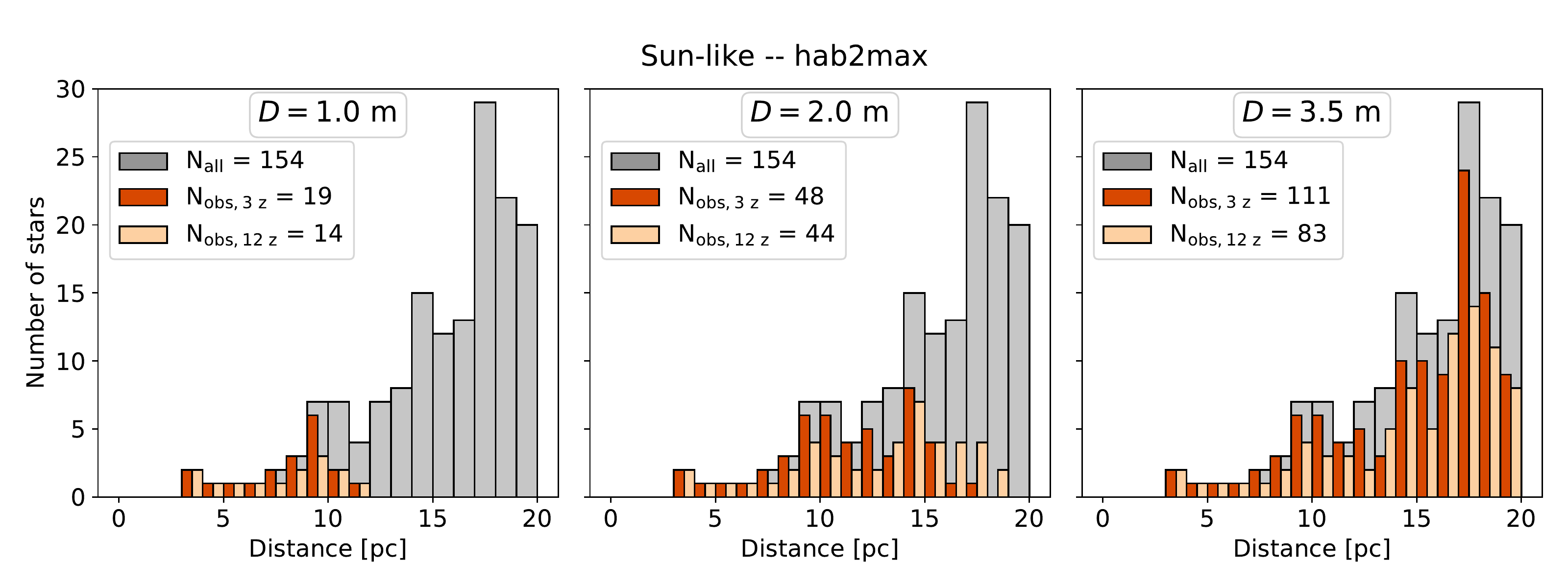}
        \caption{Observed stars as a function of the distance for the sample of Sun-like stars. Each of the six panels shows the total number of stars in gray and the number of observed stars for two different scenarios: (1) median exozodiacal dust level of 3~zodi and interferometric baseline length optimization for the center of the opt HZ in dark orange, and (2) the same, but with a median exozodiacal dust level of 12~zodi in light orange. The top row is based on the hab2min and the bottom row is based on the hab2max planet populations (see Section~\ref{sec:planet_population}), while the left, middle, and right columns show the predicted yields for 1, 2, and 3.5~m mirrors, respectively.}
        \label{fig:dist_sun}
    \end{figure*}
    
    \begin{figure*}
        \centering
        \includegraphics[width=\textwidth]{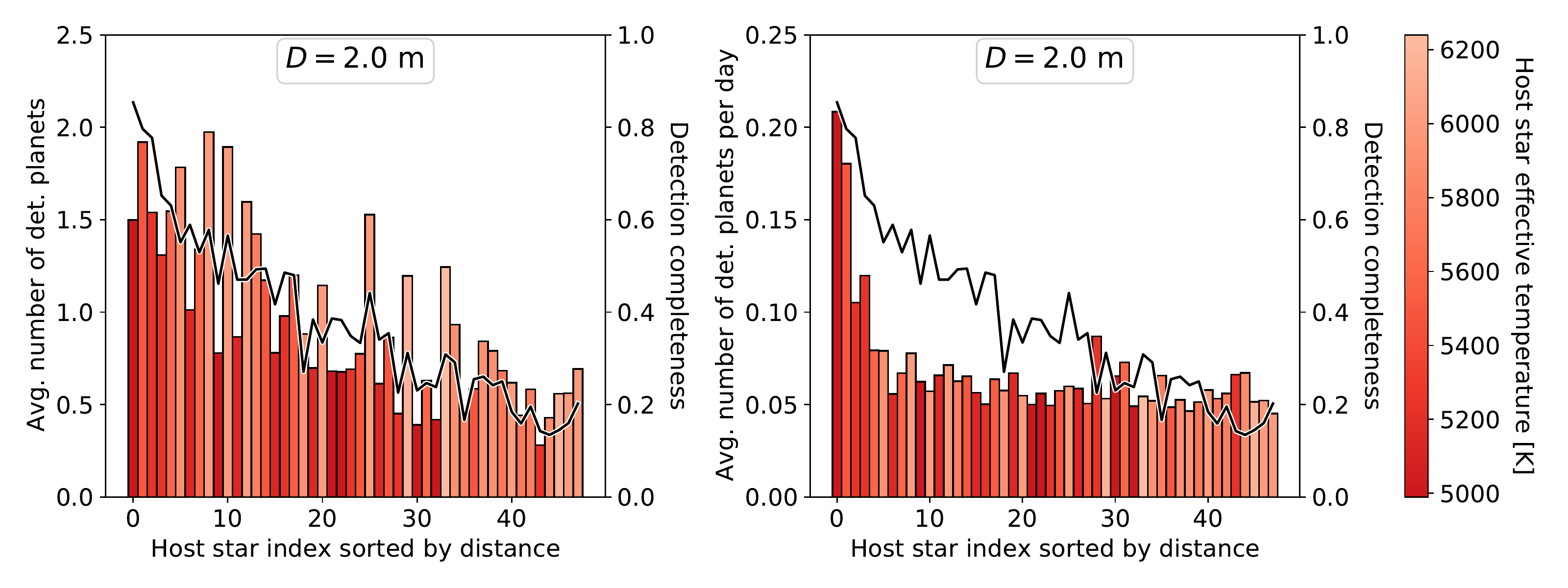}
        \caption{Predicted planet yield for each individual star. Left: Average number of detected planets for each observed Sun-like star, sorted by distance, for the hab2max planet population with a median exozodiacal dust level of 3~zodi and observed with 2~m mirrors. The spectral type is encoded in color, and the detection completeness is shown as a solid black line on the secondary y-axis. Right: Same, but for the average number of detected planets normalized by the observing time in days.}
        \label{fig:completeness}
    \end{figure*}
    
    Each of the different scenarios considered in our parameter study has its own optimized observing sequence of host stars from our stellar sample. Figure~\ref{fig:dist_sun} shows the distance distribution of Sun-like stars from our stellar sample in gray and the number of stars that is actually observed in the scenarios with a median exozodiacal dust level of 3 and 12~zodi in dark and light orange, respectively. Based on the previous findings, we only considered scenarios with the interferometric baseline length optimization for the center of the opt HZ here because the predicted yield is largely independent of this parameter. With 1~m mirrors, only stars within $\sim12~\text{pc}$ are observed. With 2~m mirrors, the distance limit increases to $\sim20~\text{pc}$ and with 3.5~m mirrors, the optimized observing sequence appears to be distance limited (because our stellar sample only contains stars out to 20~pc). It is likely that with a larger stellar sample, the predicted yield would further increase because there would be some more distant stars with a higher probability of detecting a planet that could be observed. For reference, the same plot for FGK-type stars is shown in Figure~\ref{fig:dist_fgk}. The main difference is that with 3.5~m mirrors, the observing sequence is not distance limited as strongly as for the Sun-like stars because for the FGK-type stars, more stars in the 12--20~pc regime can be observed.
    
    Studying the number of observed stars as a function of the distance reveals three findings. The two intuitive ones are that if planets are more frequent or if the median exozodiacal dust level is higher, fewer stars are observed. The first is because if planets are more frequent, it is often a more efficient use of time to remain observing a given star for slightly longer and detect a slightly fainter, but frequently occurring planet and to avoid the time penalty for slewing to another star. The second is because if the median exozodiacal dust level, and therefore the background noise level, is higher, a given star needs to be observed for longer to detect a given planet. The third and less intuitive finding is that with a higher median exozodiacal dust level, the optimizer sometimes prefers to look at more distant stars. This is especially evident in the lower middle panel of Figure~\ref{fig:dist_sun}, which shows the hab2max planet population observed with 2~m mirrors. This behavior of the optimizer can be explained by the fact that earlier-type stars have a higher planet occurrence rate (see Figure~\ref{fig:n0_teff_dependence}). An increase in the background noise level results in a preference of detecting brighter planets, and because the planet occurrence rate around earlier-type stars is higher, they are also more likely to host a brighter planet near the inner edge of the HZ. Consequently, it can be advantageous in some cases to observe a slightly more distant but earlier-type star.
    
    The fact that earlier-type stars have a higher planet occurrence rate also has a strong influence on the left panel of Figure~\ref{fig:completeness}, which shows the average number of detected planets for each observed Sun-like star, sorted by distance, for the hab2max planet population with a median exozodiacal dust level of 3~zodi and observed with 2~m mirrors. We decided to show this plot for the 2~m mirrors, because with 1~m mirrors, the predicted yield is very low, and with 3.5~m mirrors, the optimized observing sequence appears to be distance limited. The earlier-type stars shown in light red have a significantly higher predicted yield and detection completeness than the later-type stars shown in dark red. Detection completeness is here defined as the ratio of the number of planets that can be detected with \emph{LIFE} to the number of simulated planets. The higher predicted yield for earlier-type stars can be explained by longer integration times, but the fact that the optimizer prefers to observe these stars for longer is a direct consequence of their higher probability to host planets. The right panel of Figure~\ref{fig:completeness} shows the average number of detected planets normalized by the observing time, which is the quantity that is actually maximized by the optimizer. Unsurprisingly, the distribution is relatively flat beyond the nearest $\text{about five}$ stars because the optimizer observes an earlier-type star for exactly that much longer than a later-type star until their detection efficiencies (and therefore their planet yields per observing time) equalize. Last, Figure~\ref{fig:completeness} also highlights that there is a trend to observe a higher fraction of early-type stars with increasing distance. While the detection completeness naturally decreases with distance, this allows keeping the planet yield per observing time on a roughly constant level.
    
    \begin{figure*}
    \centering
    \includegraphics[width=\textwidth]{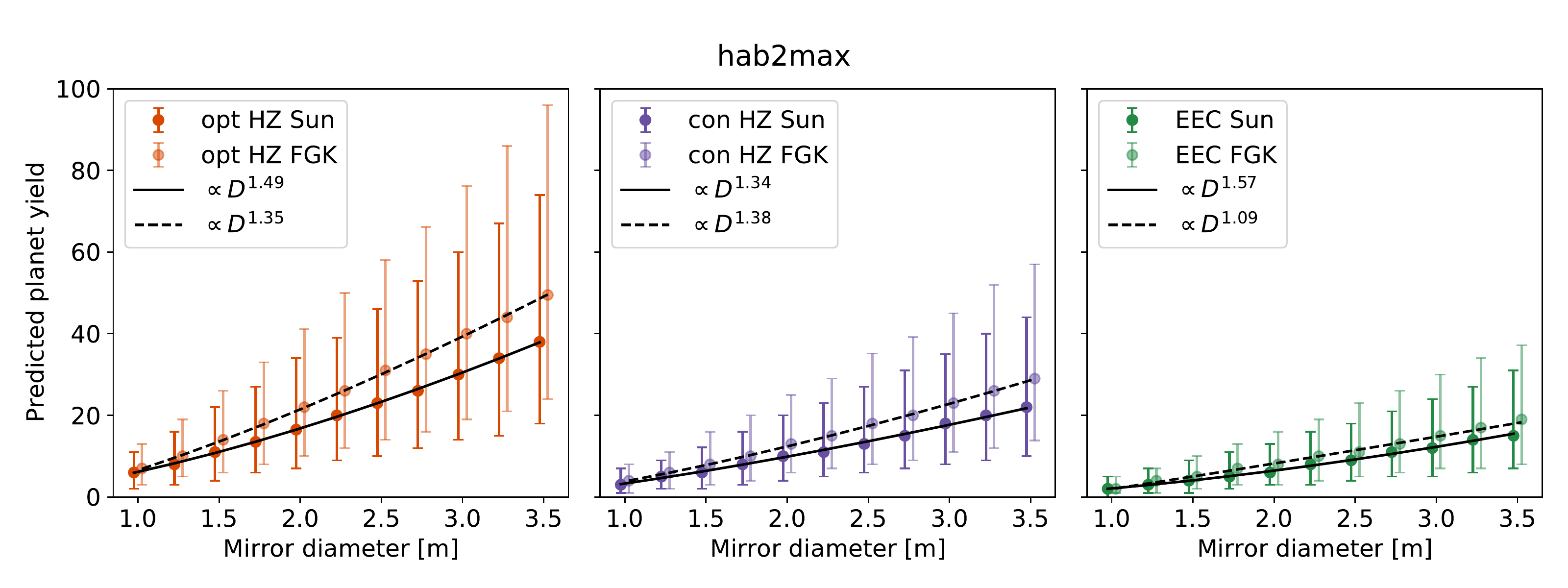}
    \caption{Predicted planet yield for the Sun-like and FGK-type stars as a function of the mirror diameter $D$ from 1~m to 3.5~m in steps of 0.25~m. We show from left to right the predicted planet yields for the opt HZ, the con HZ, and the exo-Earth candidates, for the hab2max planet population with a median exozodiacal dust level of 3~zodi. The data points for the Sun-like and FGK-type stars are slightly offset to the left and right, respectively, to increase their visibility. We overlay in solid and dashed black the power-laws fit to the data with the best-fit power-law exponent shown in the legend. The wavelength range and the total throughput are kept constant at 4--$18.5~\text{\textmu m}$ and 5\%, respectively.}
    \label{fig:impact_diam}
    \end{figure*}
    
    \begin{figure*}[ht!]
        \centering
        \includegraphics[width=\textwidth]{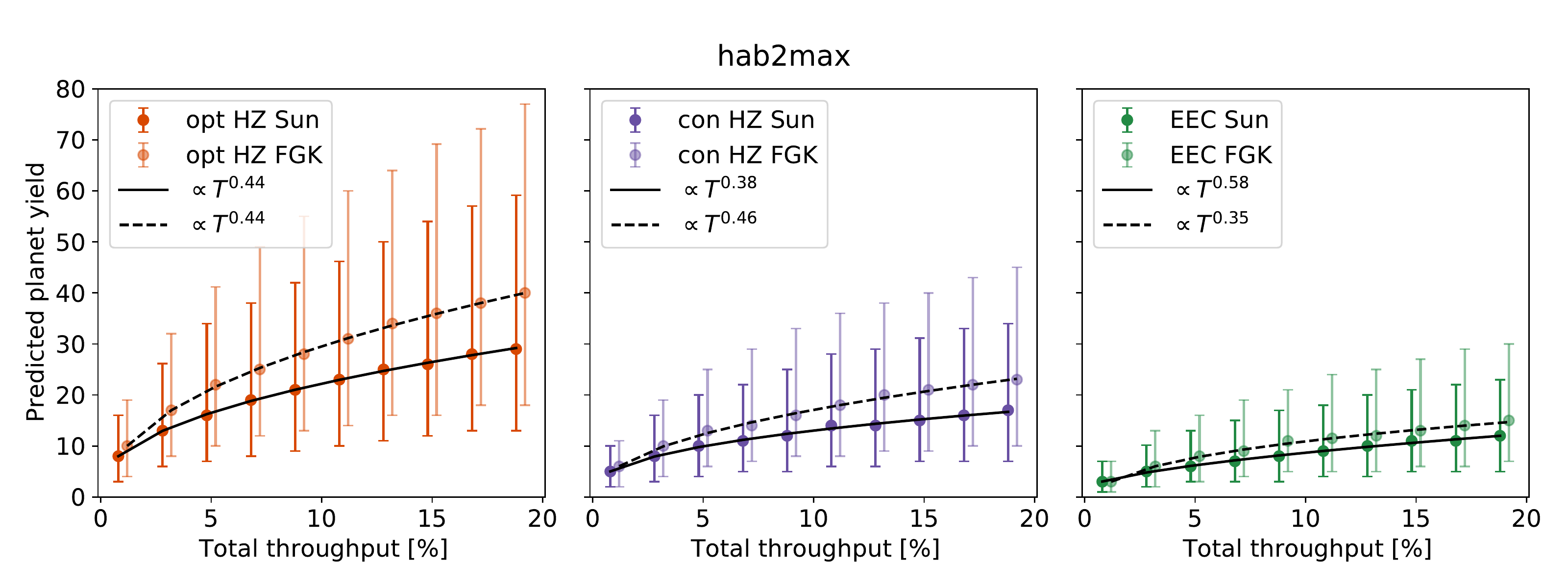}
        \includegraphics[width=\textwidth]{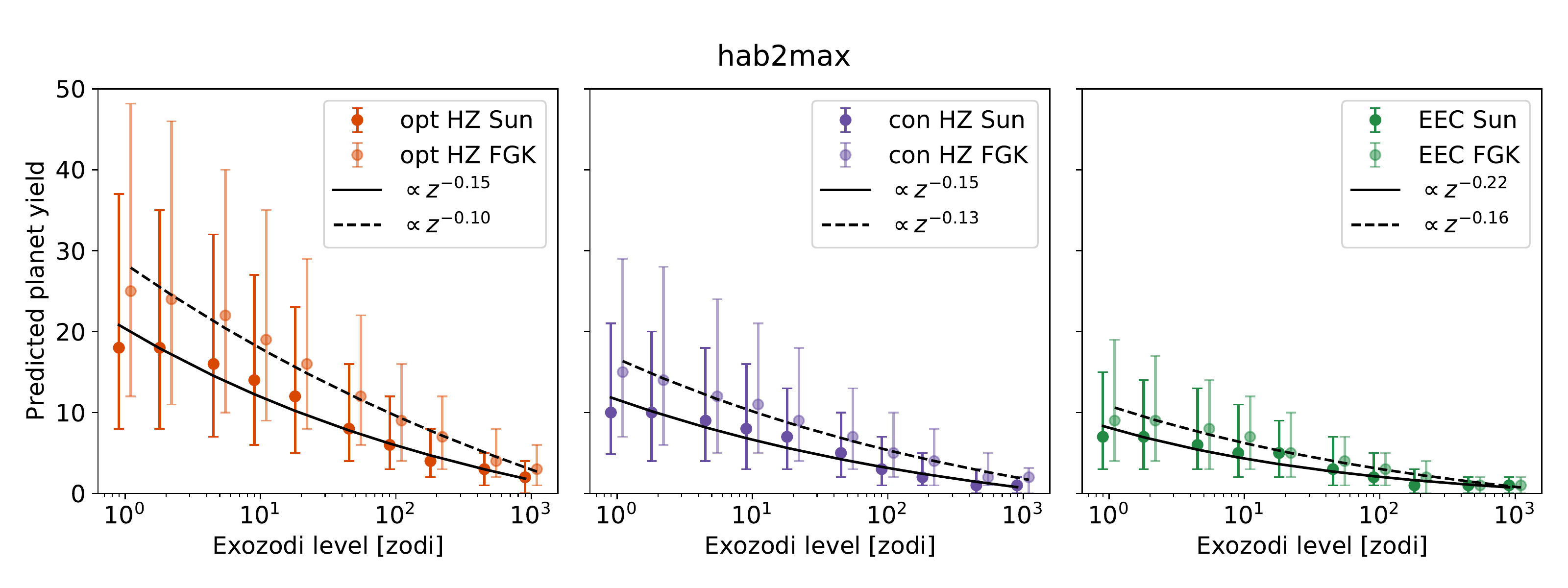}
        \includegraphics[width=\textwidth]{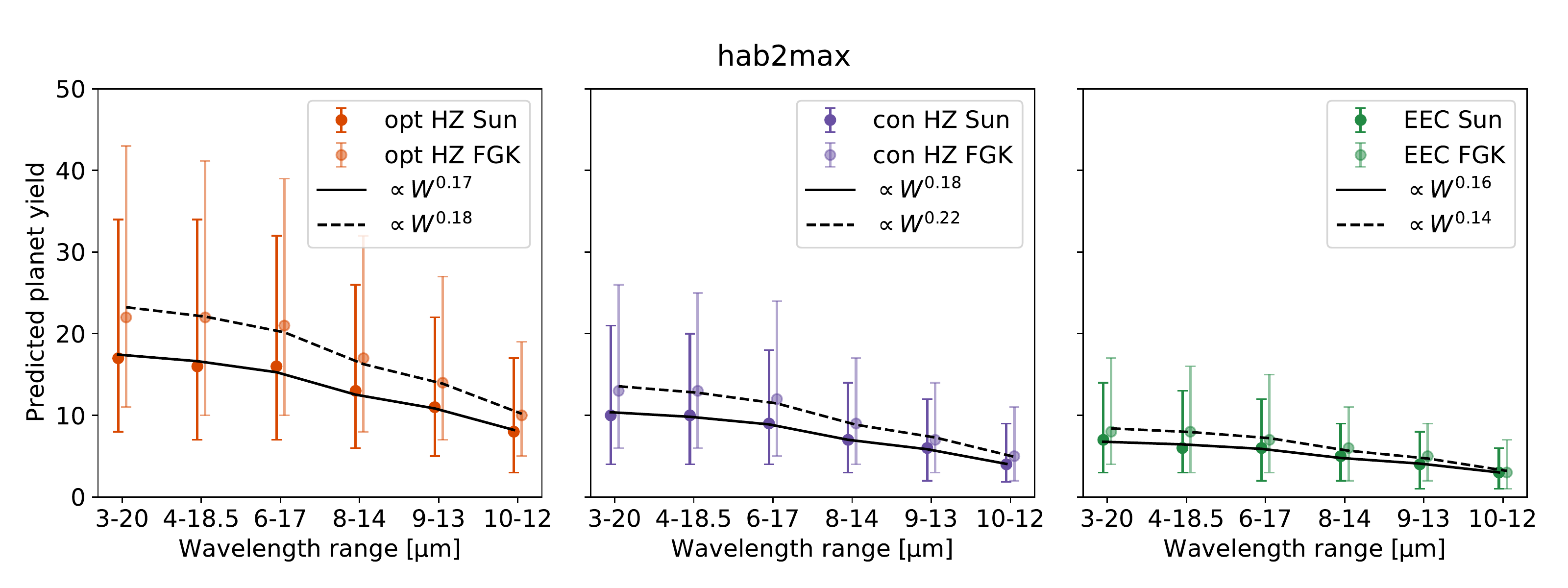}
        \caption{Same as Figure~\ref{fig:impact_diam}, but for different mission and astrophysical parameters. The mirror size is kept constant at 2~m. The top three panels show the predicted yield as a function of the total throughput $T$ from 1\% to 19\% in steps of 2\%. The middle three panels show the predicted planet yield as a function of the exozodiacal dust level $z$ from 1~zodi to 1000~zodi in logarithmic steps of $\log_{10}(\Delta z) \sim 0.3$. The bottom three panels show the predicted planet yield for different wavelength ranges. In the latter case, the overlaid power laws are functions of the width $W$ of the wavelength range, and the x-axes do not scale linearly with $W$.}
        \label{fig:impact}
    \end{figure*}
    
    \begin{figure*}
        \centering
        \includegraphics[width=\textwidth]{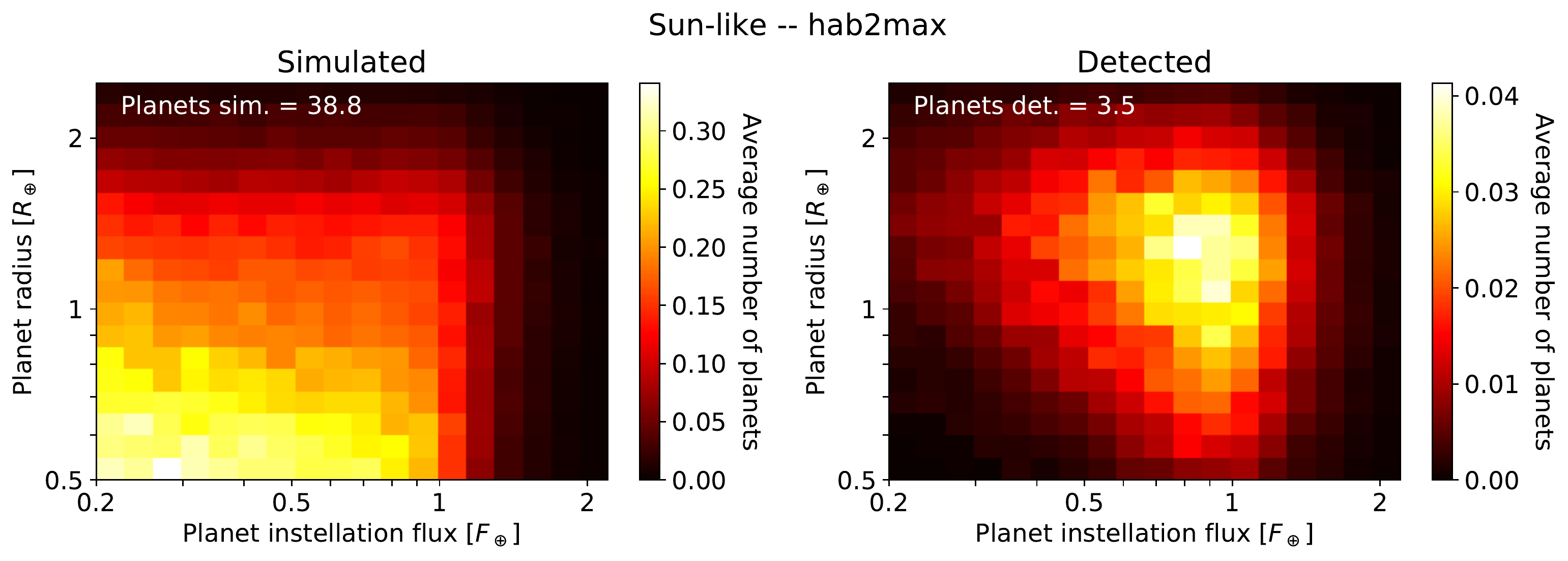}
        \caption{Average number of simulated (left) and detectable (right) habitable planets for the \emph{LIFE} mission obtained by coupling our Monte Carlo simulations with the probabilistic habitability assessment tool \texttt{HUNTER} \citep{zsom2015}. A habitable planet is here defined as a planet on which liquid surface water can exist. Both plots show the sample of Sun-like stars with the hab2max planet population and a median exozodiacal dust level of 3 zodi} observed with the 2 m mirrors, 5\% total throughput, and the interferometric baseline length optimization for the center of the opt HZ.
        \label{fig:hunter}
    \end{figure*}
    
    \section{Discussion}
    \label{sec:discussion}
    
    \subsection{Mission parameter study}
    \label{sec:mission_parameter_study}
    
    The results presented in Section~\ref{sec:results} are all based on the mission parameters shown in Table~\ref{tab:mission_parameters}. However, the predicted planet yield strongly depends on many of these parameters, and their assumed values are subject to change based on technology development and funding constraints. To enable trade-off studies and understand how changes in the different mission parameters impact the scientific potential of the \emph{LIFE} mission, we investigated the predicted planet yield as a function of the mirror diameter, the total throughput, the exozodiacal dust level, and the spectral coverage.
    
    Figure~\ref{fig:impact_diam} shows the predicted planet yield $Y$ as a function of the mirror diameter $D$ and reveals a scaling of approximately $Y \propto D^{3/2}$ for the Sun-like stars and slightly weaker for the FGK-type stars. This is a weaker relation than the $Y \propto D^{1.97}$ scaling for single-aperture reflected-light missions found by \citet{stark2015}. The main reason for the weaker scaling is that the resolution of \emph{LIFE} is controlled by the baselines between the free-flying collector spacecraft and is therefore independent of the mirror diameter $D,$ while in \citet{stark2015}, the resolution of a single-aperture reflected-light telescope scales with $\lambda/D$, where $\lambda$ is the observing wavelength. Moreover, the signal of the spatially resolved sources (mainly exozodiacal and local zodiacal light) does not depend on the mirror size to first order, further weakening the scaling from what would be expected for photon noise-limited point-source observations.
    
    Figure~\ref{fig:impact} shows the same analysis as a function of the total throughput, the exozodiacal dust level, and the spectral coverage (i.e., the wavelength range that the S/N is integrated over). The mirror size is kept constant at 2~m. For the total throughput $T$, we find a scaling of approximately $Y \propto T^{1/2}$, which agrees well with the theoretical expectation that the signal-to-noise ratio scales roughly with the square root of the observing time (higher throughput is equivalent to longer observing time in our \texttt{LIFEsim} simulations). Moreover, the scaling that we find for the total throughput of \emph{LIFE} is similar to the $Y \propto T^{0.35}$ scaling found by \citet{stark2015} for single-aperture reflected-light missions. Based on these scaling relations for the mirror diameter and the total throughput, we derive requirements of either 3~m/5\% or 2~m/20\% when we focus on Sun-like stars alone and of either 2.5~m/5\% or 2~m/10\% when we focus on all FGK-type stars for the \emph{LIFE} mission to achieve its goal of detecting 30 potentially habitable planets. For the exozodiacal dust level $z$, we find a weak scaling of approximately $Y \propto z^{-0.15}$ , which is again very similar to the $Y \propto z^{-0.17}$ scaling found by \citet{stark2015} for single-aperture reflected-light missions. Both studies make the optimistic assumption that the exozodiacal light can be subtracted down to the photon noise limit. Nevertheless, given the still large uncertainties in the median exozodiacal dust level even for Sun-like stars \citep[median 3~zodi, 95\% confidence upper limit 27~zodi, see][]{ertel2020}, this analysis illustrates that exozodiacal dust can potentially have a significant impact on the scientific potential of a future exo-Earth imaging space mission, and systems with high levels of exozodiacal dust ($\gtrsim 10~\text{zodi}$) should ideally be avoided a priori. Finally, for the width $W$ of the wavelength range, we find a similarly weak scaling as for the exozodiacal dust level, suggesting that a somewhat reduced wavelength range would be acceptable during the search phase. However, we recall that a smaller wavelength range will also decrease the precision to which the temperature and radius of a planet can already be constrained in the search phase \citep[see][]{konrad2021}, which might make the selection and prioritization of targets for the characterization phase of the mission more difficult.
    
    For completeness, we repeated the same parameter study with the hab2min planet population. The corresponding plots are shown in Figure~\ref{fig:impact_hab2min} and reveal overall similar scalings for the dependence on the mirror diameter, the total throughput, and the wavelength range. We note that there are noticeable differences in some cases, especially when the predicted planet yields are low and the curves are rather flat, leading to large uncertainties in the best-fit scaling parameters. However, only for the dependence on the median exozodi level are the scaling parameters between the hab2min and the hab2max cases consistently different, with a weaker scaling being observed with the hab2min planet population. This finding is related to the detectability of $P_\text{orb} > 500$~d planets, for which the hab2min and hab2max planet populations differ. Multiple effects such as the dependence of the exozodiacal disk size on the assumed exozodi level and the sensitivity of \emph{LIFE} as a function of the orbital separation are active, and a more detailed analysis that goes beyond the empirical scaling laws presented here is beyond the scope of this paper.
    
    \subsection{Revisits}
    \label{sec:re-visits}
    
    Observing a nearby target for a second time (or even more often frequently) to detect planets that could not be observed in the previous visit(s) can sometimes be significantly more efficient than observing a new but more distant target. Studies for single-aperture reflected-light missions conducted by \citet{stark2015} have shown that allowing for so-called revisits when optimizing an observing sequence can increase the expected number of detected HZ planets by up to $\sim40\%$. While revisits are not simulated in the current version of \texttt{LIFEsim} due to the complications illustrated in \citet{stark2015}, \cite{dannert2021} constrained the magnitude of the improvement that revisits can yield in the context of \emph{LIFE}. By moving all planets in the sample to the position on their orbit where they are observed with maximum S/N, 18\% is predicted to be an upper bound for the improvement achievable with revisits. To predict a lower bound for the improvement, the search phase is divided into a blind search phase (as described in this work) and a follow-up phase. In the follow-up phase, planets observed in the blind search phase with insufficient S/N to be detected are moved to the optimal observing position on their orbit. Then, this optimized sample is observed with the remaining observing time, yielding a 6\% lower bound for the improvement possible with revisits. This approach additionally predicts that revisits are efficient in increasing the completeness of the observed systems. Compared to a single-visit observing sequence, revisits can increase the completeness by up to 30\% without reducing the total number of detected planets. This can be helpful when studying the architecture of the detected planetary systems, for instance.
    
    The difference of a factor of $\sim2$ in the improvement from revisits between \emph{LIFE} and reflected-light missions arises because the thermal emission observations are independent of the planet phase angle. For reflected-light observations, planets are more easily detected in gibbous phase than in crescent phase. Statistically speaking, at a given moment, 50\% of the planets are in crescent phase and might remain undetected even if they could be resolved from their host star. However, when simulating revisits, these planets eventually move into gibbous phase and will be detected. For thermal emission observations, the planets will already be detected during the first visit, regardless of their phase. This accordingly decreases the improvement that can be achieved by revisiting the system.
    
    \subsection{Probabilistic habitable zone}
    \label{sec:probabilistic_habitable_zone}
    
    The discrete nature of the HZs defined in Table~\ref{tab:habitable_zones} and considered throughout this work thus far limits their usefulness for quantifying the number of actually habitable planets (i.e., the number of planets on which liquid surface water can exist) because only a fraction of the planets residing within these HZs will ultimately be habitable. For our own Solar System, this habitability fraction is 50\% for both the opt HZ and con HZ (see Figure~\ref{fig:habitable_zones}). We note that this fraction might have changed throughout the evolution of the Solar System \citep{tuchow2021}. To estimate the number of actually habitable planets orbiting Sun-like stars that the \emph{LIFE} mission could detect, we coupled our Monte Carlo simulations with the \texttt{HUNTER} tool\footnote{\url{https://github.com/andraszsom/HUNTER}} from \citet{zsom2015}. \texttt{HUNTER} considers a grid of exoplanet atmospheres with multiple parameters that can impact the habitability of a planet (e.g., atmospheric composition, relative humidity, or surface pressure) to assess the probability that a planet receiving a given instellation flux and having a given radius can harbor liquid surface water. In the current implementation, the priors for the surface pressure and the relative humidity are uniform in log-space, and the priors for the surface albedo, the N${}_2$ mixing ratio, and the CO${}_2$ mixing ratio are uniform. By multiplying the number of simulated or detectable planets from our Monte Carlo simulations with the habitability probability from \texttt{HUNTER}, we obtain an estimate for the number of simulated or detectable planets that are actually habitable.
    
    Figure~\ref{fig:hunter} shows the average number of simulated and detectable habitable planets for the \emph{LIFE} mission obtained with the default settings for Sun-like stars from \texttt{HUNTER}. As in the previous sections, Figure~\ref{fig:hunter} considers the sample of Sun-like stars with the hab2max planet population and a median exozodiacal dust level of 3 zodi observed with 2~m mirrors and 5\% total throughput. On average, $\sim3.5$ out of the $\sim38.8$ simulated and habitable planets can be detected around Sun-like stars. It appears that most of the habitable planets according to \texttt{HUNTER} have small radii and instellation fluxes and cannot be detected by \emph{LIFE}. Plots for the sample of FGK-type stars and the hab2min planet population can be found in Appendix~\ref{app:hunter}. For FGK-type stars for instance, there are $\sim4.5$ detections out of the $\sim76.2$ simulated and habitable planets. This result suggests a higher detection rate of habitable planets for Sun-like stars ($\sim9\%$) than for the broader sample of FGK-type stars ($\sim6\%$). However, while the planet occurrence rate model from \citet{bryson2021} covers the entire parameter space of EECs, the probabilistic nature of the \texttt{HUNTER} tool means that even planets outside the parameter space simulated here could be habitable. This is especially the case for slightly less illuminated and smaller planets \citep[see Figure~5 of][]{zsom2015}. Hence, the numbers presented here should be considered with caution and might be slightly different if a broader space of planet parameters were simulated. We finally recall that the numbers obtained from coupling our Monte Carlo simulations with the \texttt{HUNTER} tool have similarly high relative uncertainties as the numbers presented throughout this work thus far, and they also depend on the mission parameters in a similar fashion.
    
    \subsection{Combining separately optimized observing sequences}
    \label{sec:combining_separately_optimized_observing_sequences}
    
    We showed in Section~\ref{sec:mission_parameter_study} that the predicted planet yield scales roughly with the square root of the total throughput (see Figure~\ref{fig:impact}). To first order, the total throughput is directly proportional to the duration of the search phase, which we have assumed to be 2.5~years (see Table~\ref{tab:mission_parameters}). Hence, the predicted planet yield also scales roughly with the square root of the duration of the search phase, and if the search phase were cut to half of its time (1.25~years), significantly more than half of the predicted planet yield for the full 2.5~years would still be obtained. This raises the interesting possibility of combining multiple separately optimized observing sequences by reducing the duration assigned to each of these observing sequences while implicitly increasing their observing efficiency. For example, if the same scaling that we find here for Sun-like and FGK-type stars also holds for M dwarfs, the \emph{LIFE} mission could search for planets orbiting Sun-like or FGK-type stars for half of the search phase (1.25~years) and for planets orbiting M dwarfs for the other half of the search phase. Then, for each of these two 50\% chunks of the search phase, we would still obtain $\sim71\%$ of the predicted planet yield for the full 2.5~years.
    
    \section{Summary and conclusions}
    \label{sec:summary_and_conclusions}
    
    Previous studies have predominantly found that a space-based mid-infrared nulling interferometer mission searching for rocky planets within the HZ of their host star is strongly biased toward detections around M dwarfs \citep{kammerer2018,quanz2018,quanz2021}. This stems from the fact that such an interferometric mission is mainly sensitivity limited when searching for rocky HZ planets, and M dwarfs are much more abundant in the solar neighborhood than FGK-type stars. A common misconception is that HZ planets would also be more frequent around M dwarfs than around Sun-like stars, but the results from \citet{bryson2021} showed that the HZ occurrence rates for Sun-like stars are at least the same as those for M dwarfs, if not higher (see also Table~\ref{tab:habitable_zones}). Because the habitability prospects for planets orbiting M dwarfs are still unclear, we focused here on the question of how many rocky HZ planets a space-based mid-infrared nulling interferometer mission such as \emph{LIFE} could detect around Sun-like (effective temperature within 4800--6300~K) and FGK-type stars (effective temperature within 3940--7220~K). For this purpose, we presented new Monte Carlo simulations of the search phase of such a mission, effectively predicting its expected planet yield. Our simulations are based on the planet occurrence rate model from \citet{bryson2021}, specifically inferred for Sun-like stars, and also propagate the uncertainties in this occurrence rate model, the orbital configurations, and the exozodiacal dust levels from \citet{ertel2020} to the final results.
    
    We find that an interferometric array consisting of four 2~m telescopes operating in the $4$--$18.5~\text{\textmu m}$ wavelength range with a total throughput of 5\% and targeting Sun-like stars is expected to find $10^{+10}_{-5}$ and $16^{+18}_{-10}$ planets within the opt HZ defined according to \citet{kopparapu2014}, assuming a pessimistic and optimistic extrapolation of the \emph{Kepler} planet occurrence rates beyond 500~days, respectively. If targeting FGK-type stars, these numbers increase to $16^{+12}_{-8}$ and $22^{+20}_{-12}$, respectively. The expected number of detected exo-Earth candidates, which are planets with radii between 0.8--1.4 Earth radii (lower bound depends on spectral type, see Table~\ref{tab:habitable_zones}) and instellation fluxes between 0.356--1.107 solar constants (upper and lower bounds depend on spectral type, see Table~\ref{tab:habitable_zones}), is approximately one-third of the expected number of detected planets within the opt HZ. In Section~\ref{sec:results}, we presented the predicted planet yield for 24 different scenarios featuring different  extrapolation schemes for the planet occurrence rate, exozodiacal dust levels, interferometric baseline length optimization schemes, and instrument parameters. Furthermore, in Section~\ref{sec:mission_parameter_study}, we investigated the dependence of the predicted planet yield on the mirror diameter, the total throughput, the exozodiacal dust level, and the spectral coverage in more detail. Overall, we observe no significant dependence of the predicted planet yield $Y$ on the interferometric baseline length optimization scheme, while the mirror diameter $D$ has the strongest impact (approximately $Y \propto D^{3/2}$), followed by the total throughput $T$ (approximately $Y \propto T^{1/2}$), and finally the exozodiacal dust level $z$ (approximately $Y \propto z^{-0.15}$) and the width of the accessible wavelength range with a similarly weak scaling. Finally, we used the probabilistic habitability assessment tool \texttt{HUNTER} to predict that roughly half of the detected exo-Earth candidates might indeed be habitable with conditions that would allow liquid surface water to exist. Finding these habitable planets and constraining their frequency is one of the main objectives of the \emph{LIFE} mission.
    
    This new work clearly shows that the \emph{LIFE} mission would be able to detect a statistically significant sample of Earth-like exoplanets around Sun-like stars. Although focusing on these planets is more difficult, mainly because Sun-like stars are less frequent in the solar neighborhood than M dwarfs, the \emph{LIFE} mission could achieve its goal of detecting 30 potentially habitable planets during a search phase of 2.5~years when focusing solely on Sun-like stars assuming $\sim3~\text{m}$ mirrors and a throughput of 5\%. This result assumes an optimistic extrapolation of the \emph{Kepler} planet occurrence rates (the hab2max model). With smaller $2~\text{m}$ mirrors, a total throughput of $\sim20\%$ would result in a similar planet yield. If the stellar sample were  extended to all FGK-type stars, the goal of 30 potentially habitable planets could already be achieved with either a 2.5~m mirror and 5\% total throughput combination or a 2~m mirror and 10\% total throughput combination. These planets would then be prime targets for an in-depth characterization during a second mission phase. Constructing an observing sequence that only partially investigates Sun-like (or more broadly, FGK-type) stars while leaving some time to search for HZ planets around M dwarfs would further reduce the requirements on mirror diameter and total throughput, although this option hinges on the unclear prospects for habitability around M dwarfs.
    
    \begin{acknowledgements}
    The authors would like to thank Steve Bryson for sharing the parameter posterior of their planet occurrence rate model. The authors would further like to thank Christopher~C. Stark, Denis Defr\'ere, Olivier Absil, and Romain Laugier for valuable feedback and discussions that helped to improve the manuscript. This work has been carried out within the framework of the National Centre of Competence in Research PlanetS supported by the Swiss National Science Foundation. SPQ acknowledges the financial support of the SNSF. This research has made use of the SIMBAD database, operated at CDS, Strasbourg, France. This research has made use of the Washington Double Star Catalog maintained at the U.S. Naval Observatory. The manuscript was substantially improved following helpful comments from an anonymous referee.
    \end{acknowledgements}
    
    \bibliographystyle{aa}
    \bibliography{bibliography_new}
    
    \begin{appendix}
    \onecolumn
    \section{Predicted planet yield}
    \label{app:predicted_planet_yield}
    
    \begin{table}[h!]
    \caption{Predicted planet yield around Sun-like stars for the different planet populations, median exozodiacal dust levels $\Tilde{z}$, interferometric baseline length optimization schemes, and mirror diameters $D$.}
    \label{tab:planet_yield}
    \centering
    \begin{tabular}{cccc|cccc}
    \hline\hline
    Population & $\Tilde{z}$ (zodis) & baseline optimization & $D$ (m) & $N_\text{det}$ & $N_\text{det,opt HZ}$ & $N_\text{det,con HZ}$ & $N_\text{det,EEC}$\\
    \hline
    \multirow{6}{*}{hab2min} & \multirow{6}{*}{3} & \multirow{3}{*}{opt HZ} & 1.0 & $10.0^{+6.0}_{-4.0}$ & $3.0^{+4.0}_{-2.0}$ & $2.0^{+2.0}_{-2.0}$ & $1.0^{+2.0}_{-1.0}$\\
    & & & 2.0 & $30.0^{+16.0}_{-10.0}$ & $11.0^{+10.0}_{-6.0}$ & $5.0^{+6.0}_{-3.0}$ & $4.0^{+3.0}_{-3.0}$\\
    & & & 3.5 & $70.0^{+37.0}_{-21.2}$ & $27.0^{+22.0}_{-14.0}$ & $14.0^{+13.0}_{-7.0}$ & $10.0^{+9.0}_{-5.0}$\\
    & & \multirow{3}{*}{con HZ} & 1.0 & $8.0^{+6.0}_{-4.0}$ & $3.0^{+4.0}_{-2.0}$ & $2.0^{+2.0}_{-2.0}$ & $1.0^{+2.0}_{-1.0}$\\
    & & & 2.0 & $26.0^{+15.0}_{-9.0}$ & $10.0^{+10.0}_{-5.0}$ & $6.0^{+5.0}_{-4.0}$ & $4.0^{+4.0}_{-3.0}$\\
    & & & 3.5 & $65.0^{+36.0}_{-22.0}$ & $26.0^{+21.0}_{-13.0}$ & $14.0^{+13.0}_{-7.0}$ & $10.0^{+9.0}_{-6.0}$\\
    \hline
    \multirow{6}{*}{hab2max} & \multirow{6}{*}{3} & \multirow{3}{*}{opt HZ} & 1.0 & $11.0^{+7.0}_{-5.0}$ & $4.0^{+5.0}_{-3.0}$ & $2.0^{+3.0}_{-2.0}$ & $1.0^{+3.0}_{-1.0}$\\
    & & & 2.0 & $39.0^{+22.0}_{-15.0}$ & $14.0^{+14.0}_{-8.0}$ & $8.0^{+8.0}_{-5.0}$ & $6.0^{+6.0}_{-4.0}$\\
    & & & 3.5 & $92.0^{+55.0}_{-34.2}$ & $36.0^{+36.0}_{-19.0}$ & $22.0^{+20.0}_{-12.2}$ & $15.0^{+16.0}_{-8.0}$\\
    & & \multirow{3}{*}{con HZ} & 1.0 & $10.0^{+6.0}_{-5.0}$ & $4.0^{+4.0}_{-3.0}$ & $2.0^{+3.0}_{-2.0}$ & $1.0^{+2.2}_{-1.0}$\\
    & & & 2.0 & $35.0^{+21.0}_{-14.0}$ & $13.0^{+14.0}_{-7.0}$ & $8.0^{+9.0}_{-5.0}$ & $6.0^{+6.0}_{-4.0}$\\
    & & & 3.5 & $85.0^{+55.2}_{-33.0}$ & $35.0^{+34.2}_{-19.0}$ & $21.5^{+22.5}_{-12.5}$ & $15.0^{+16.0}_{-9.0}$\\
    \hline
    \multirow{6}{*}{hab2min} & \multirow{6}{*}{12} & \multirow{3}{*}{opt HZ} & 1.0 & $7.5^{+4.5}_{-3.5}$ & $2.0^{+3.0}_{-1.0}$ & $1.0^{+2.0}_{-1.0}$ & $1.0^{+1.0}_{-1.0}$\\
    & & & 2.0 & $25.0^{+11.0}_{-9.0}$ & $8.0^{+8.0}_{-4.0}$ & $4.0^{+4.0}_{-2.0}$ & $3.0^{+3.0}_{-2.0}$\\
    & & & 3.5 & $54.0^{+28.2}_{-18.0}$ & $20.0^{+18.0}_{-10.0}$ & $11.0^{+11.0}_{-6.0}$ & $7.0^{+8.0}_{-4.0}$\\
    & & \multirow{3}{*}{con HZ} & 1.0 & $6.0^{+5.0}_{-3.0}$ & $2.0^{+3.0}_{-1.0}$ & $1.0^{+2.0}_{-1.0}$ & $1.0^{+1.0}_{-1.0}$\\
    & & & 2.0 & $21.0^{+11.0}_{-8.0}$ & $8.0^{+7.0}_{-5.0}$ & $4.0^{+5.0}_{-3.0}$ & $3.0^{+3.0}_{-2.0}$\\
    & & & 3.5 & $50.0^{+27.0}_{-17.0}$ & $20.0^{+17.0}_{-10.0}$ & $11.0^{+11.0}_{-6.0}$ & $7.0^{+8.0}_{-4.0}$\\
    \hline
    \multirow{6}{*}{hab2max} & \multirow{6}{*}{12} & \multirow{3}{*}{opt HZ} & 1.0 & $11.0^{+6.0}_{-6.0}$ & $4.0^{+4.0}_{-3.0}$ & $2.0^{+3.0}_{-2.0}$ & $1.0^{+2.0}_{-1.0}$\\
    & & & 2.0 & $35.0^{+20.0}_{-13.0}$ & $13.0^{+13.0}_{-8.0}$ & $7.0^{+8.0}_{-4.0}$ & $5.0^{+6.0}_{-3.0}$\\
    & & & 3.5 & $81.0^{+48.0}_{-32.0}$ & $33.0^{+31.0}_{-18.0}$ & $19.0^{+19.0}_{-10.0}$ & $14.0^{+14.0}_{-8.0}$\\
    & & \multirow{3}{*}{con HZ} & 1.0 & $9.0^{+6.0}_{-4.0}$ & $3.0^{+5.0}_{-2.0}$ & $2.0^{+3.0}_{-2.0}$ & $1.0^{+2.0}_{-1.0}$\\
    & & & 2.0 & $30.0^{+19.0}_{-12.0}$ & $12.0^{+12.0}_{-7.0}$ & $7.0^{+8.0}_{-4.0}$ & $5.0^{+6.0}_{-3.0}$\\
    & & & 3.5 & $75.0^{+46.2}_{-30.2}$ & $32.0^{+30.0}_{-17.0}$ & $19.0^{+20.0}_{-11.0}$ & $14.0^{+13.0}_{-9.0}$\\
    \hline
    \multicolumn{8}{l}{\textbf{Notes.} \parbox[t]{14.0 cm}{HZ denotes the habitable zone for whose center the interferometric baseline length was optimized.}}
    \end{tabular}
    \end{table}
    
    \newpage
    
    \section{Observed stars with the FGK-type sample}
    \label{app:fgk-type_stars}
    
    \FloatBarrier
    
    \begin{figure*}[h!]
        \centering
        \includegraphics[width=\textwidth]{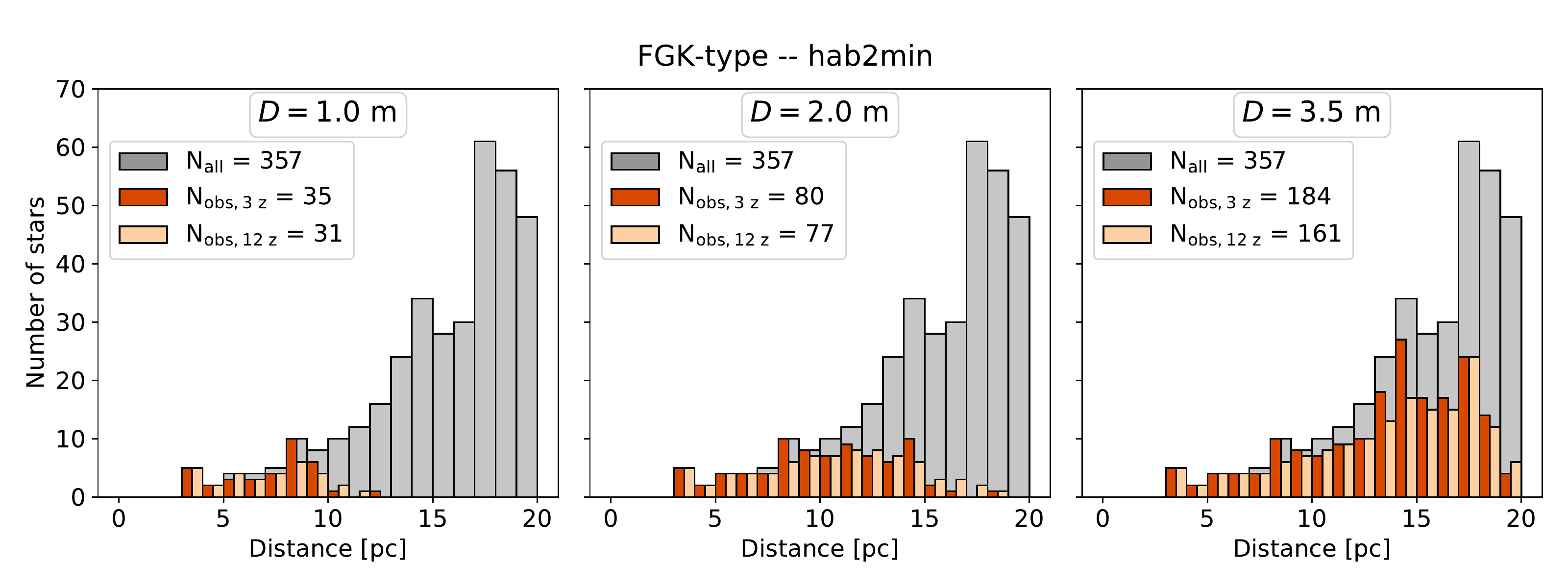}
        \includegraphics[width=\textwidth]{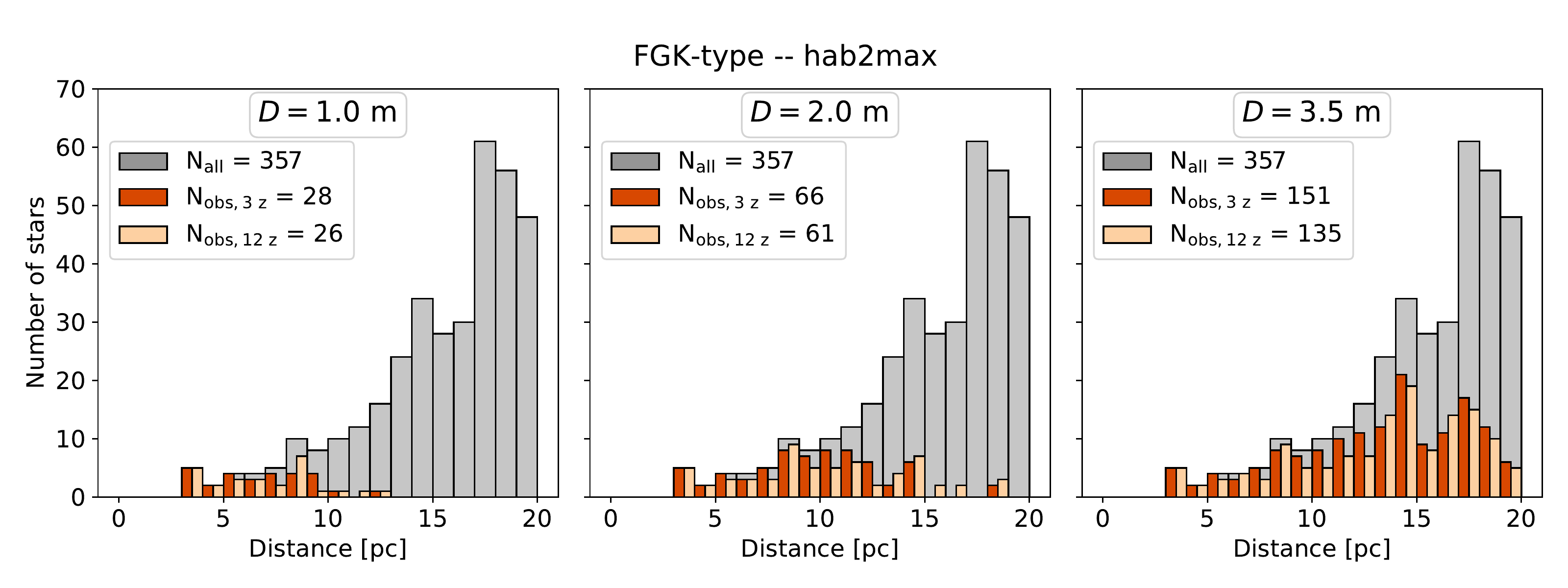}
        \caption{Same as Figure~\ref{fig:dist_sun}, but for the sample of FGK-type stars.}
        \label{fig:dist_fgk}
    \end{figure*}
    
    \newpage
    
    \section{Mission parameter study with the hab2min planet population}
    \label{app:mission_parameter_study}
    
    \begin{figure*}[h!]
        \centering
        \includegraphics[width=0.825\textwidth]{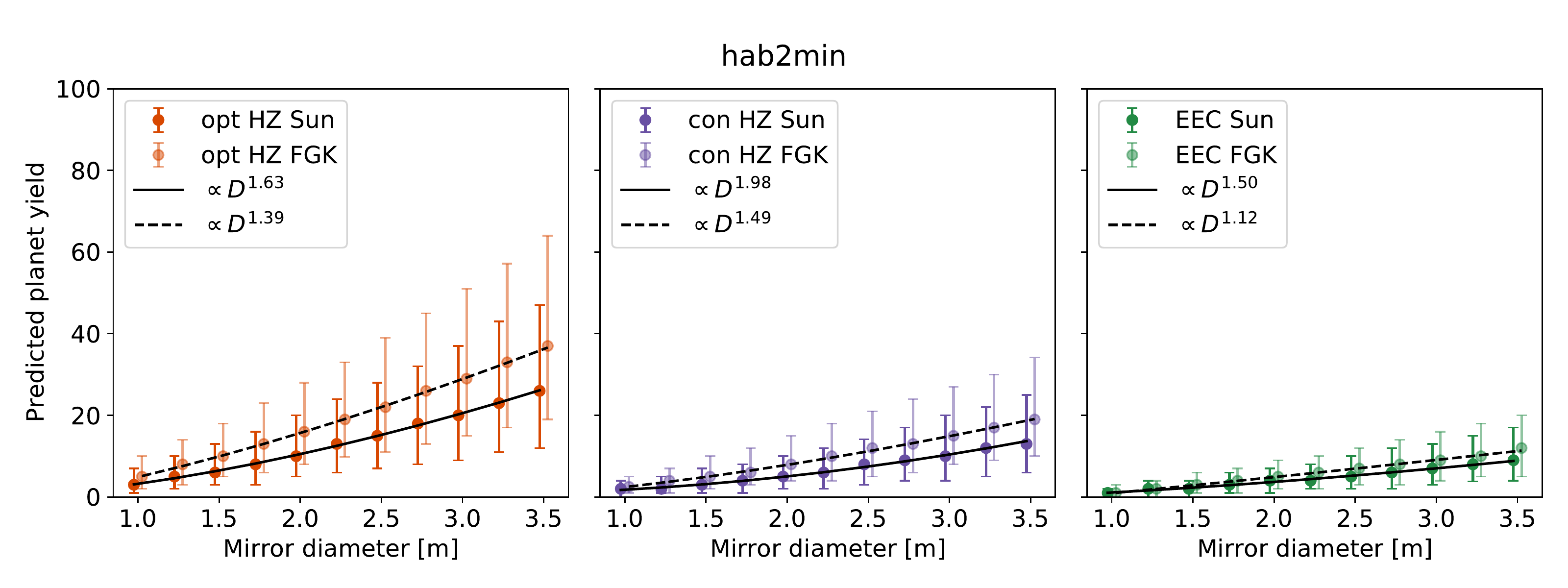}
        \includegraphics[width=0.825\textwidth]{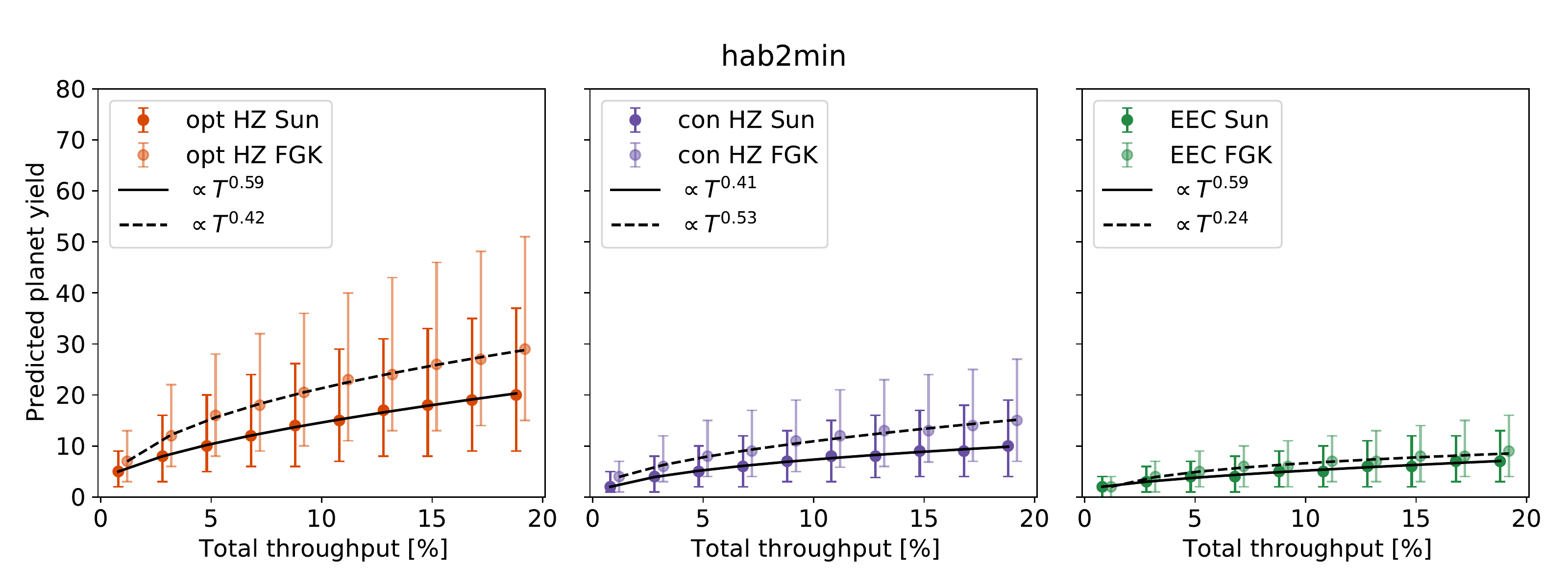}
        \includegraphics[width=0.825\textwidth]{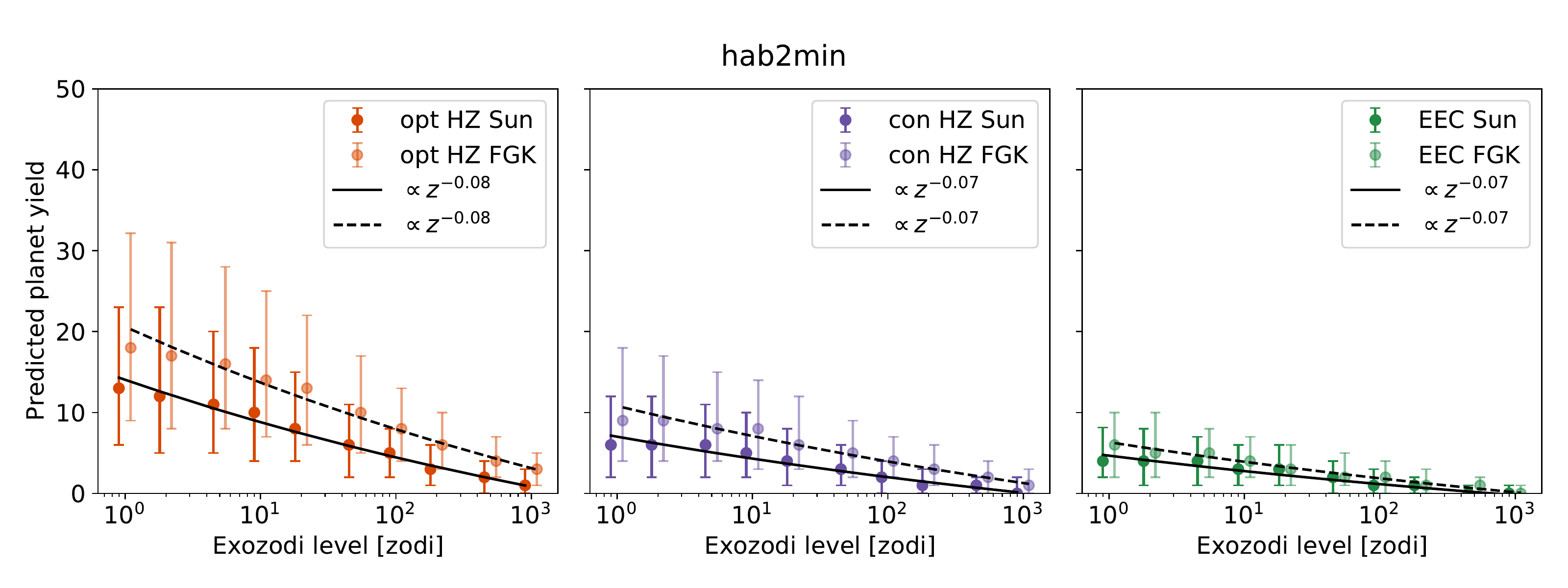}
        \includegraphics[width=0.825\textwidth]{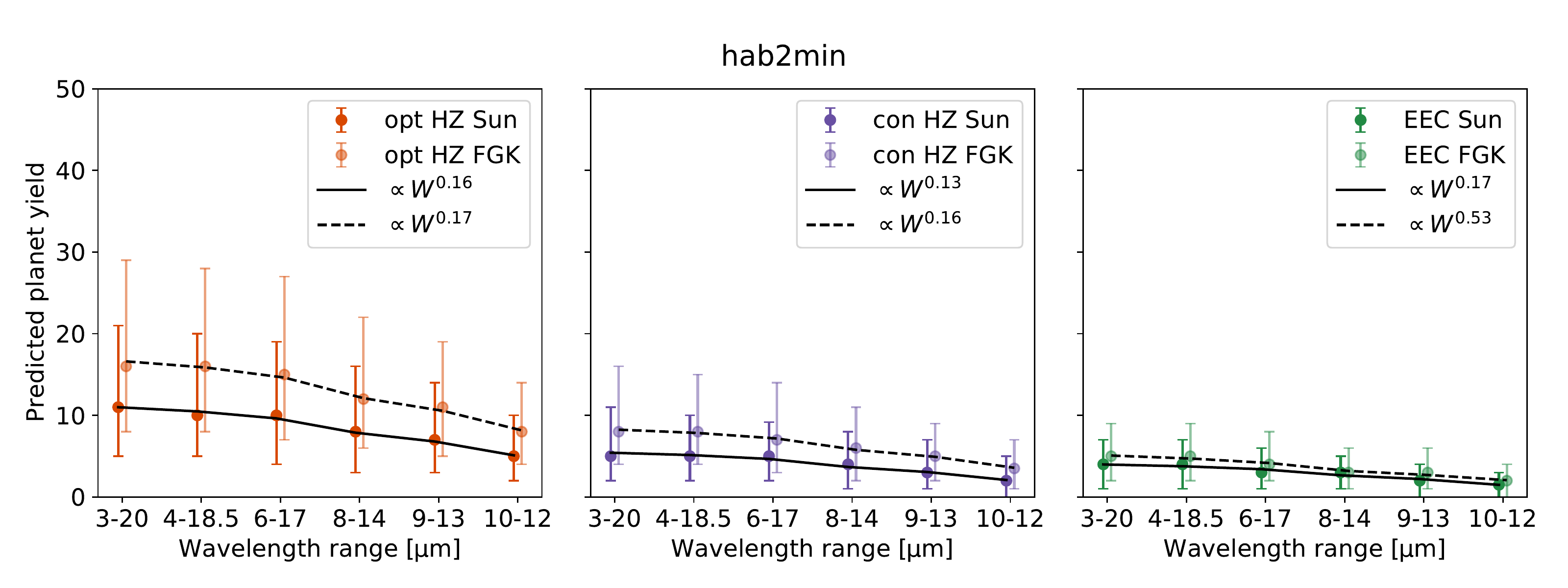}
        \caption{Same as Figure~\ref{fig:impact_diam} and~\ref{fig:impact}, but for the hab2min planet population.}
        \label{fig:impact_hab2min}
    \end{figure*}
    
    \newpage
    
    \section{HUNTER plots for the FGK-type sample and the hab2min planet population}
    \label{app:hunter}
    
    \begin{figure*}[h!]
        \centering
        \includegraphics[width=0.9\textwidth]{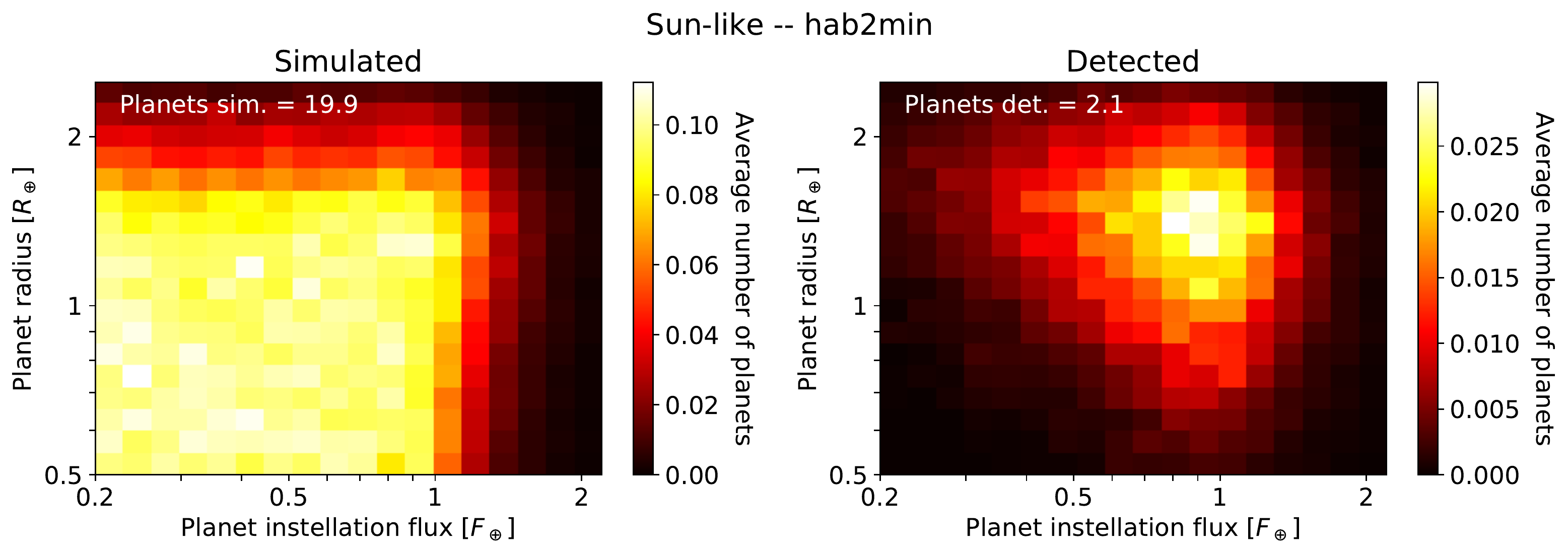}
        \includegraphics[width=0.9\textwidth]{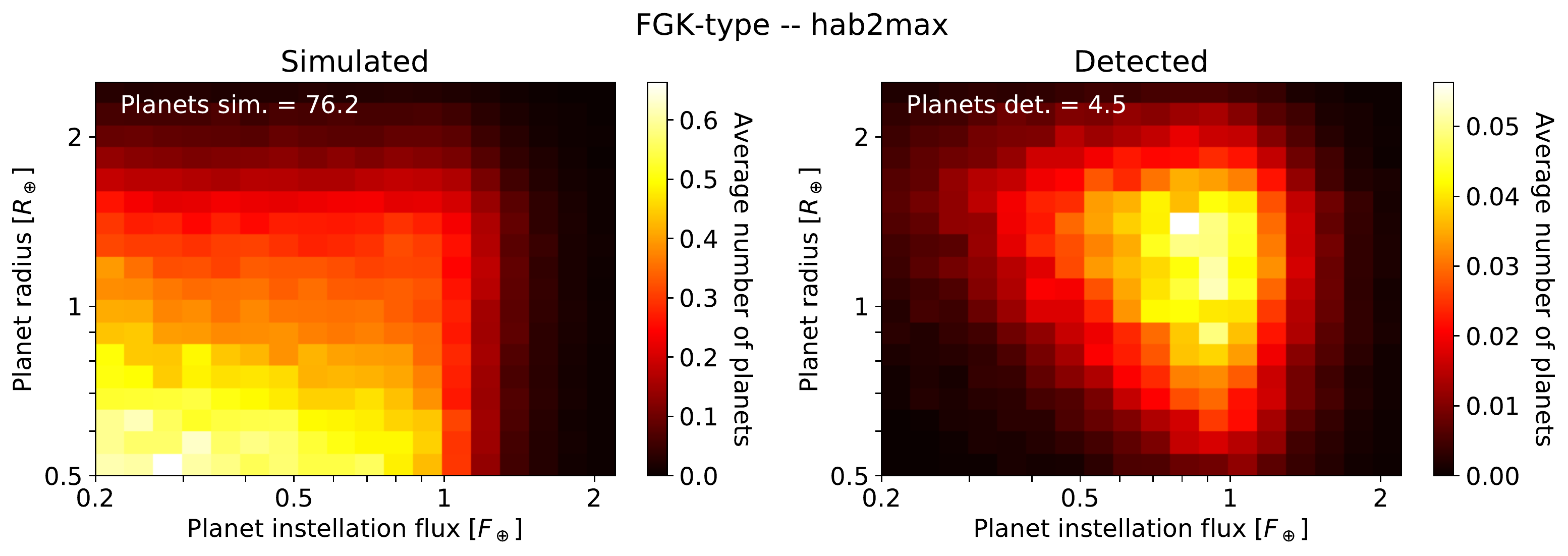}
        \includegraphics[width=0.9\textwidth]{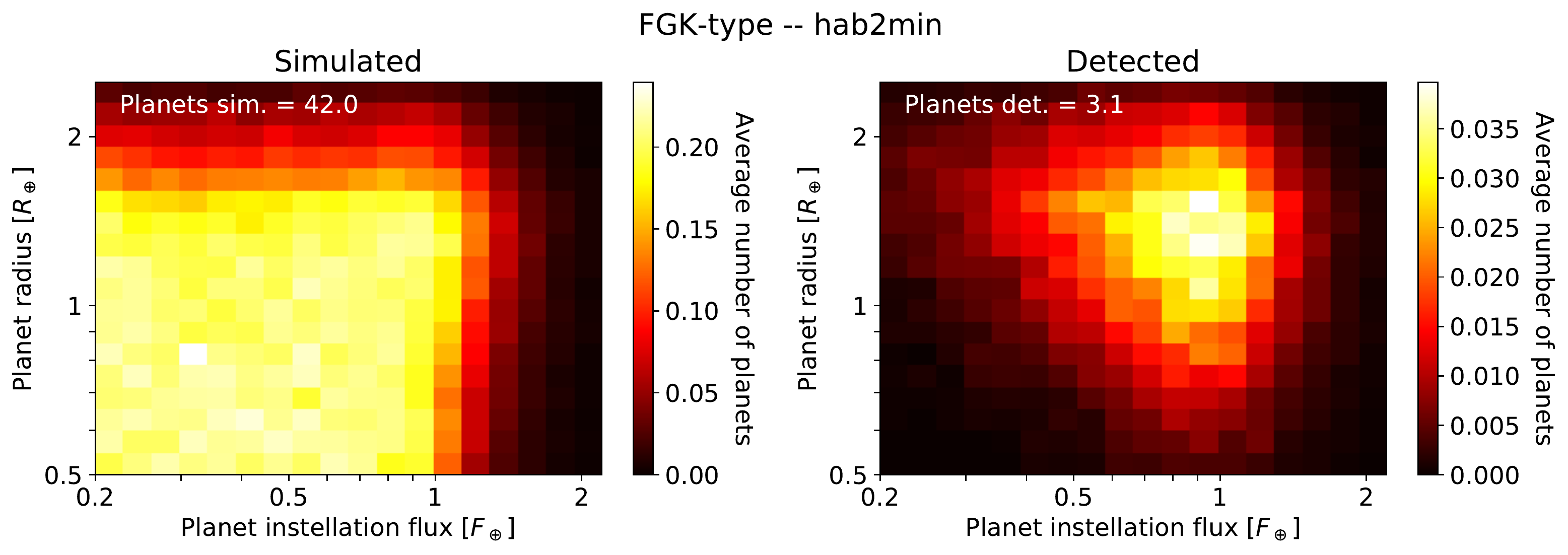}
        \caption{Same as Figure~\ref{fig:hunter}, but for Sun-like stars with the hab2min planet population (top panels), for FGK-type stars with the hab2max planet population (middle panels), and for FGK-type stars with the hab2min planet population (bottom panels).}
        \label{fig:hunter_hab2min}
    \end{figure*}
    \end{appendix}

\end{document}